\documentclass{jaa}
\usepackage{natbib}
\usepackage[dvipsnames]{xcolor}
\definecolor{xlinkcolor}{cmyk}{1,1,0,0}

 \usepackage[bookmarks=false,         
     pdfnewwindow=true,      
     colorlinks=true,    
     linkcolor=xlinkcolor,     
     citecolor=xlinkcolor,     
     filecolor=xlinkcolor,  
     urlcolor=xlinkcolor,      
final=true
 ]{hyperref}

\usepackage{graphicx}
\usepackage{lastpage}
\usepackage{multicol}
\usepackage{aas_macros}
\usepackage{subcaption}
\usepackage{rotating}
\usepackage{multirow}

\newcommand{\degr}{\ensuremath{^\circ}}
\newcommand{\asat}{{\em AstroSat}}
\newcommand{\fermi}{{\em Fermi}}
\newcommand{\kw}{{\em Konus-Wind}}

\newcommand{\sw}[1]{\texttt{#1}}
\newcommand{\geant}{\sw{GEANT4}}

\hypersetup{colorlinks=true,allcolors={cyan}}
\newcommand{\todo}[1]{\textit{\color{brown} #1}}

\begin{document}\sloppy

\title{The AstroSat Mass Model: Imaging and Flux studies of off-axis sources with CZTI}

\author{\href{https://orcid.org/0000-0001-5536-4635}{Sujay Mate}\textsuperscript{1,2},
\href{https://orcid.org/0000-0001-9856-1866}{Tanmoy Chattopadhyay}\textsuperscript{3,4}, 
\href{https://orcid.org/0000-0002-6112-7609}{Varun Bhalerao}\textsuperscript{1,*},
\href{https://orcid.org/0000-0002-7729-1964}{E. Aarthy}\textsuperscript{5}
\href{https://orcid.org/0000-0003-0477-7645}{Arvind Balasubramanian}\textsuperscript{1,6},
\href{https://orcid.org/0000-0003-3352-3142}{Dipankar Bhattacharya}\textsuperscript{7},
\href{https://orcid.org/0000-0001-6621-259X}{Soumya Gupta}\textsuperscript{7},
\href{}{Krishnan Kutty}\textsuperscript{8},
\href{https://orcid.org/0000-0003-3431-6110}{N.P.S. Mithun}\textsuperscript{5}, 
\href{https://orcid.org/0000-0003-2932-3666}{Sourav Palit}\textsuperscript{1},
\href{https://orcid.org/0000-0003-0833-0533}{A. R. Rao}\textsuperscript{8},
\href{https://orcid.org/0000-0001-6332-1723}{Divita Saraogi}\textsuperscript{1},
\href{https://orcid.org/0000-0002-2050-0913}{Santosh Vadawale}\textsuperscript{5},
\and \href{https://orcid.org/0000-0003-1501-6972}{Ajay Vibhute}\textsuperscript{7}
}

\affilOne{\textsuperscript{1} Indian Institute of Technology Bombay, Mumbai, India\\}
\affilTwo{\textsuperscript{2} IRAP, Universit\'e de Toulouse, CNES, CNRS, UPS, Toulouse, France\\}
\affilThree{\textsuperscript{3} Department of Physics, Stanford University, 382 Via Pueblo Mall, Stanford CA 94305, USA\\}
\affilFour{\textsuperscript{4} Kavli Institute of Astrophysics and Cosmology, 452 Lomita Mall,
Stanford, CA 94305, USA\\}
\affilFive{\textsuperscript{5} Physical Research Laboratory, Ahmedabad, Gujarat, India\\}
\affilSix{\textsuperscript{6} Department of Physics and Astronomy, Texas Tech University, Box 1051, Lubbock, TX 79409-1051, USA\\}
\affilSeven{\textsuperscript{7} The Inter-University Centre for Astronomy and Astrophysics, Pune, India\\}
\affilEight{\textsuperscript{8} Tata Institute of Fundamental Research, Mumbai, India\\}

\twocolumn[{

\maketitle

\corres{varunb@iitb.ac.in}

\msinfo{---}{---}

\begin{abstract}

The Cadmium Zinc Telluride Imager (CZTI) on AstroSat is a hard X-ray coded-aperture mask instrument with a primary field of view of 4.6 x 4.6 degrees (FWHM). The instrument collimators become increasingly transparent at energies above $\sim$100 keV, making CZTI sensitive to radiation from the entire sky. While this has enabled CZTI to detect a large number of off-axis transient sources, calculating the source flux or spectrum requires knowledge of the direction and energy dependent attenuation of the radiation incident upon the detector. Here, we present a GEANT4-based mass model of CZTI and AstroSat that can be used to simulate the satellite response to the incident radiation, and to calculate an effective ``response file'' for converting the source counts into fluxes and spectra. We provide details of the geometry and interaction physics, and validate the model by comparing the simulations of imaging and flux studies with observations. Spectroscopic validation of the mass model is discussed in a companion paper,~\citet{Chattopadhyay2021}.

\end{abstract}


}] 


\doinum{12.3456/s78910-011-012-3}
\artcitid{\#\#\#\#}
\volnum{000}
\year{0000}
\pgrange{1--}
\setcounter{page}{1}
\lp{\pageref{LastPage}}

\section{Introduction}\label{sec:intro}

The Cadmium Zinc Telluride Imager on board \asat\ is a hard X-ray (20 -- 200~keV) coded aperture mask instrument with a $4.6\degr\times4.6\degr$ field of view~\citep{2014SPIE.9144E..1SS,czti}. The primary objectives of the instrument are spectroscopy, imaging, and timing studies of hard X-ray sources. At energies above $\sim100$~keV, the instrument collimators become increasingly transparent to the radiation from off-axis directions~\citep{Rao2017a}. Thus, the CZT detectors are sensitive to the entire sky up to $\sim200$~keV. This sensitivity is extended to $\sim650$~keV by the CsI anti-coincidence ``veto'' detectors installed in the instrument. This all-sky sensitivity has been leveraged for detection of transient sources like Gamma Ray Bursts (GRBs).

CZTI detected its first GRB, GRB~151006A, on the very first day it was switched on~\citep{2015GCN..18422...1B, 2016ApJ...833...86R}. In the five years since, CZTI has detected more than 400 GRBs, 88 of which are being reported for the first time in~\citet{sharma2020search}. The sensitivity of CZTI is comparable to several other GRB missions~\citep{bkb+17}, which has led to many significant results. The detection of GRB~170105A, which was missed by all major missions, conclusively proved that an ``orphan afterglow'' discovered in optical was not related to the binary black hole merger GW170104~\citep{2017GCN..20387...1M, 2017GCN..20389...1S, 2017GCN..20412...1B}. When CZTI did not detect GW170817, we inferred that the source was occulted by the Earth --- narrowing down the source localisation by a factor of two~\citep{kns+17}. The off-axis sensitivity of CZTI has also been leveraged for Earth-occultation studies of sources~\citep{Singhal2021}. Compton scattering within the CZT detectors is sensitive to polarisation of incoming photons, and has been successfully used to measure polarisation of GRBs~\citep{2019ApJ...874...70C,2019ApJ...884..123C} and the Crab pulsar~\citep{2018NatAs...2...50V}.

While CZTI has had great success in detecting off-axis transients, the interpretation of detected signals requires a detailed modelling of the instrument and satellite. Incoming X-ray photons interact with various satellite elements undergoing absorption, coherent and incoherent scattering, etc. Photons can get absorbed and re-emitted as fluorescence lines. Such interactions modify the energy, direction, and position of interaction of incoming radiation. These effects are strongly direction-dependent, based on the mass distribution of various materials in the satellite. In this paper, we present the \asat\ Mass Model: a numerical simulation of such interactions to calculate the observed spatial and energy distribution of photons for any given astrophysical source. 

The basic concepts of this mass model were introduced in \citep{2019ApJ...884..123C}. In this paper, we give the full details of the mass model and give results on imaging and flux studies. A companion paper~\citep[][hereafter Paper II]{Chattopadhyay2021} discusses details of sub-MeV spectroscopy using the mass model, and compares the results with various other sub-MeV spectroscopic methods.

Numerical simulations using the mass model are utilised for three key calculations. First, these simulations are used for mapping the observed count rates and spectra to a source spectrum. We discuss examples with count rates in this paper and spectra are discussed in Paper II. The same technique is used to calculate flux upper limits for non-detections, based on an assumed source spectrum~\citep{sharma2020search}. Secondly, mass model simulations form a key part of the measurement of polarisation of astrophysical sources~\citep{2014ExA....37..555C, 2015A&A...578A..73V}. Thirdly, the observed distribution of photons on the detectors (Detector Plane Histogram, DPH) is strongly dependent on the incident direction of the source photons. Crudely speaking, different satellite elements cast unique shadows on the detector plane, allowing us to use the entire satellite as a mask for locating source positions. A basic ray-tracing version of this concept was used to localise GRB~170105A~\citep{bkb+17}, and the more refined mass model is expected to improve such localisations.

This paper is organised as follows: In \S\ref{sec:massmodel} we describe the framework used for the numerical simulations. A detailed discussion of the model components and simplifying approximations is presented in \S\ref{sec:geometry}, it is followed by a discussion of the model physics in \S\ref{sec:phys}. In \S\ref{sec:validation}, we discuss the efficacy of the model by comparing simulations to observations. We conclude by discussing future work in \S\ref{sec:conclusion}.


\section{Numerical modelling}\label{sec:massmodel}

As discussed above, our goal is to create a numerical model that can simulate the interaction of incoming radiation with the satellite, to yield the final energies (spectrum) and positions (DPH) of photons incident on the detector. The first step is to create a detailed digital representation of the the satellite. Then, we need a software to simulate the interaction and passage of radiation through the satellite. These interactions are probabilistic in nature: multiple photons entering the satellite at exactly the same point from the same direction may all undergo different interactions with different satellite elements. The final spectrum and DPH can hence be interpreted only in the average sense, and this caveat underscores all comparisons with observations.

\asat, like any other satellite, is a complex structure and modelling all interactions is a non-trivial task. We tackle this problem with the aid of the \geant\ toolkit for particle, photons and matter interactions\footnote{\url{http://geant4.web.cern.ch/geant4/}}. \geant\ has wide range of applications in high energy physics, space sciences, and medical science~\citep{2003NIMPA.506..250A, 2006ITNS...53..270A, 2016NIMPA.835..186A}. It is an easy to use open source simulation toolkit which provides all the necessary building blocks to simulate complex particle matter interactions. \geant\ features pre-defined geometry classes and a large materials database for easy construction of elaborate geometrical structures. In addition, computer-aided design (CAD) models can also be imported into the code. The toolkit supports a wide range of physical processes for photons and particles, and allows tracking and extraction of photon properties at any stage of simulation. Lastly, it is a highly scalable toolkit making it easy to develop and test the mass model simulation on personal computers, and run it on high performance computing clusters for speed and performance. Thanks to these features, we selected \geant\ for creating the \asat\ mass model. We now discuss the satellite geometry construction in \S\ref{sec:geometry}, followed by the interaction physics in \S\ref{sec:phys}.


\section{Geometry Constructions} \label{sec:geometry}
\begin{figure}[ht!]
\includegraphics[width=0.5\textwidth]{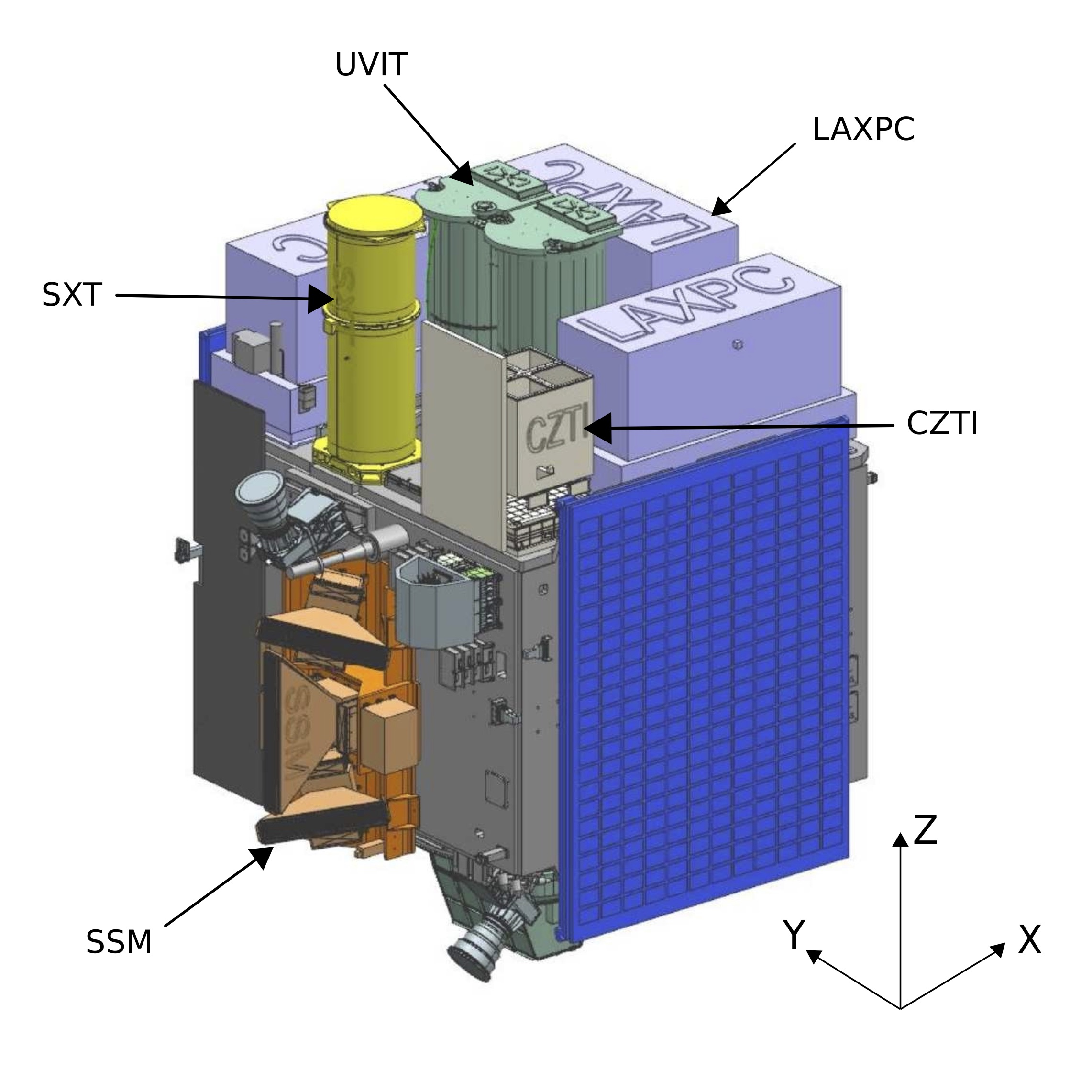}
\caption{Representative CAD model of \asat\ with solar panels folded. \label{fig:astrosat}}
\end{figure}

The basic structure of \asat\ is a cuboid of dimensions $\sim\ 2 \times 1.8 \times 1.8$~m that supports all instruments (Figure~\ref{fig:astrosat}). CZTI (\S\ref{sec:czti}) is mounted on the ``top'' deck of this cuboid, along with the Soft X-ray Telescope (SXT, \S\ref{sec:sxt}) and all three units of the Large Area X-ray Proportional Counter (LAXPC, \S\ref{sec:laxpc}). The Ultraviolet Imaging Telescope (UVIT, \S\ref{sec:uvit}) is mounted on the bottom deck and penetrates the entire satellite body. The Scanning Sky Monitor (SSM, \S\ref{sec:ssm}) is mounted on a rotating platform on one of the sides of the cuboid. Various other satellite bus components are also contained in the cuboid.

The geometry and materials of \asat\ are simulated by creating a custom detector construction class derived from the standard \geant\ \sw{G4VUserDetectorConstruction}. The geometrical shapes of the model are defined either using the built-in \geant\ geometry shapes (\sw{G4Tubes, G4Box, G4Sphere} etc.) or by importing CAD files (in case of complex geometry) in \sw{.stl} format using the \sw{CADmesh}\footnote{\url{https://github.com/christopherpoole/CADMesh}} library~\citep{Poole2012}. The material properties, both for pure and composite materials, are defined using the \sw{G4Material} class. \geant\ derives these properties from the NIST\footnote{\url{http://physics.nist.gov/PhysRefData/Star/Text/method.html}} database.

Several small-scale structural intricacies of \asat\ are unimportant for the key goals of our numerical simulations, but add a significant computational overhead in the simulations. Hence, we make certain simplifying assumptions while modelling them in \geant. In several cases, a component is replaced with a uniform block of the same size, mass, and average composition. Some other small components like Titanium screws have been completely ignored in the mass model. Thus, the total simulated mass is comparable to the actual satellite mass (Table~\ref{tab:mass}). We now discuss the specifics about the model and approximations used in each instrument:

\begin{table}
\caption{Comparison of simulated and actual masses of key components of \asat.}
\begin{tabular}{|p{1in}|c|c|}
\hline
\textbf{Instrument} & \textbf{Simulated Mass} & \textbf{Actual Mass} \\
\textbf{} & \textbf{ kg } & \textbf{ kg } \\
\hline
CZTI &  41.87 & 50.29\\
SXT  &  44.03 &  57.64\\
UVIT &  212.58 &  202.06\\
LAXPC & 350.87 &  389.10\\
SSM  &  71.53 &  71.53\\
Satellite bus and electronics & 657.22 & 668.47\\
\hline
Total & 1378.10 & 1439.09\\
\hline
\end{tabular}
\label{tab:mass}
\end{table}

\subsection{Cadmium Zinc Telluride Imager}\label{sec:czti}
\begin{figure}[ht!]
\begin{subfigure}{\columnwidth}
    \centering
    \includegraphics[width=\columnwidth]{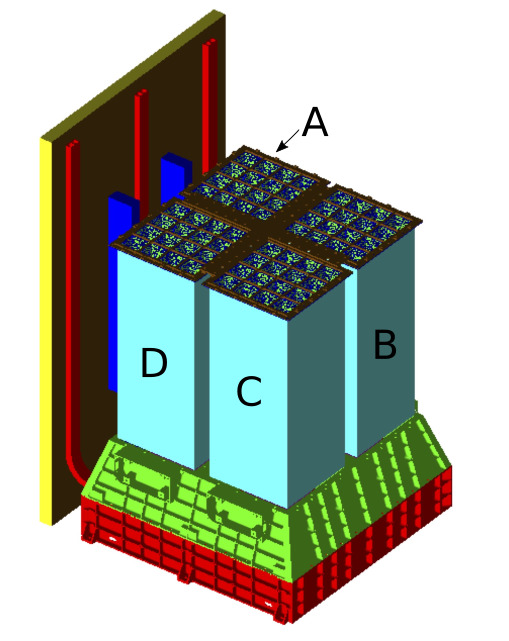}
    \caption{The entire CZT imager}
    \label{fig:czti1}
\end{subfigure}
\begin{subfigure}{\columnwidth}
    \centering
    \includegraphics[width=\columnwidth]{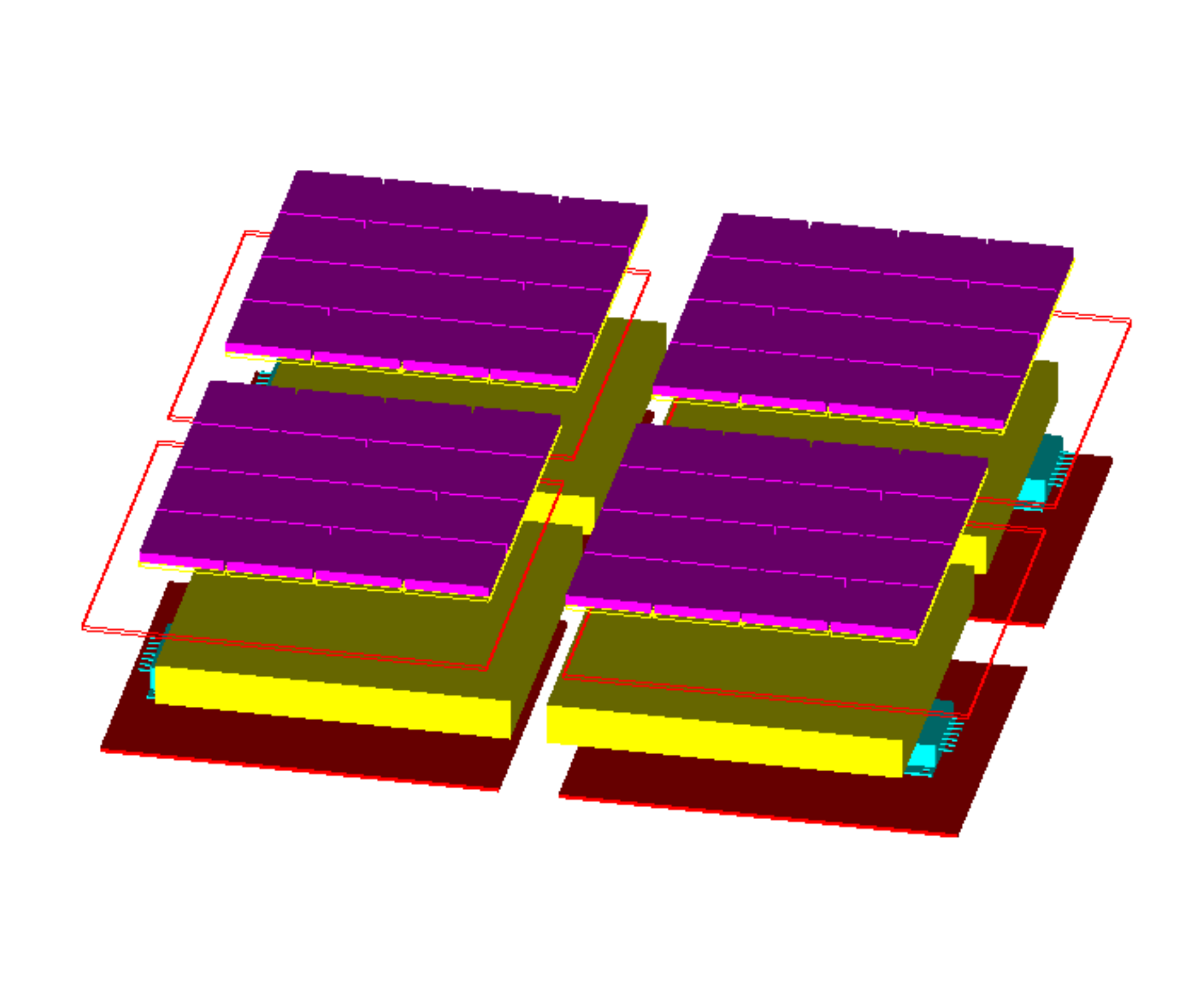}
    \caption{Relative placement of CZT and Veto detectors.}
    \label{fig:czti2}
\end{subfigure}
\caption{Panel~\subref{fig:czti1}: Rendering of the CZTI mass model. The green part houses the CZT and Veto detectors. The quadrants are labeled to show their relative orientation with respect to the radiator plate. Panel~\subref{fig:czti2}: Inner view of the green section, showing the four arrays of CZT detectors (purple) and the four Veto detectors (yellow). \label{fig:czti}}
\end{figure}

The collimators, housing, and other structural elements of CZTI are closest to the detectors, and hence have the most influence on the spatial and energy redistribution of incident photons. Hence the CZTI geometry is modelled as accurately as possible (Figure~\ref{fig:czti}). The overall construction of CZTI is discussed in \citet{bkb+17}. Some elements of the geometry are coded using the \geant\ geometry classes while the complex structure like coded masks (both top and side), heat pipes, detector and collimator housing, etc. are imported from the CAD geometry file. The coded masks are pure Tantalum while the collimators are 1 mm Aluminium with 0.07 mm Tantalum pasted on one side.

For simplicity, each CZT module has been modelled to be 40 $\times$ 40 $\times$ 5 mm instead of 39.06 $\times$ 39.06 $\times$ 5 mm. This change increases the effect area of each module by about 5\%. Therefore each pixel size in simulation is 2.5 mm instead of the actual sizes of 2.46~mm for central pixels and 2.31~mm for edge pixels. The detector composition is 43\% Cadmium, 2.8\% Zinc and 54.2\% Tellurium. To model the electronics cards of CZTI, a PCB material is defined using \sw{G4Material} class by assuming elemental composition of a regular PCB. The dimensional accuracy of cards is within 5\%.

The veto detector is defined using \geant\ geometry class as a continuous slab of CsI(Tl) with dimensions 160 $\times$ 160 $\times$ 20 mm. The composition of CsI(Tl) is 51\% Cs, 48\% I, 1\% Tl. The veto casing is imported using the CAD model. The Photo-multiplier Tube (PMT) inside the casing is not modelled. The veto electronics cards are defined similarly as in case of CZTI.

\subsection{Large Area X-ray Proportional Counter}\label{sec:laxpc}
\begin{figure*}[h]
\centering
\begin{subfigure}{0.25\textwidth}
    \centering
    \includegraphics[height=4in]{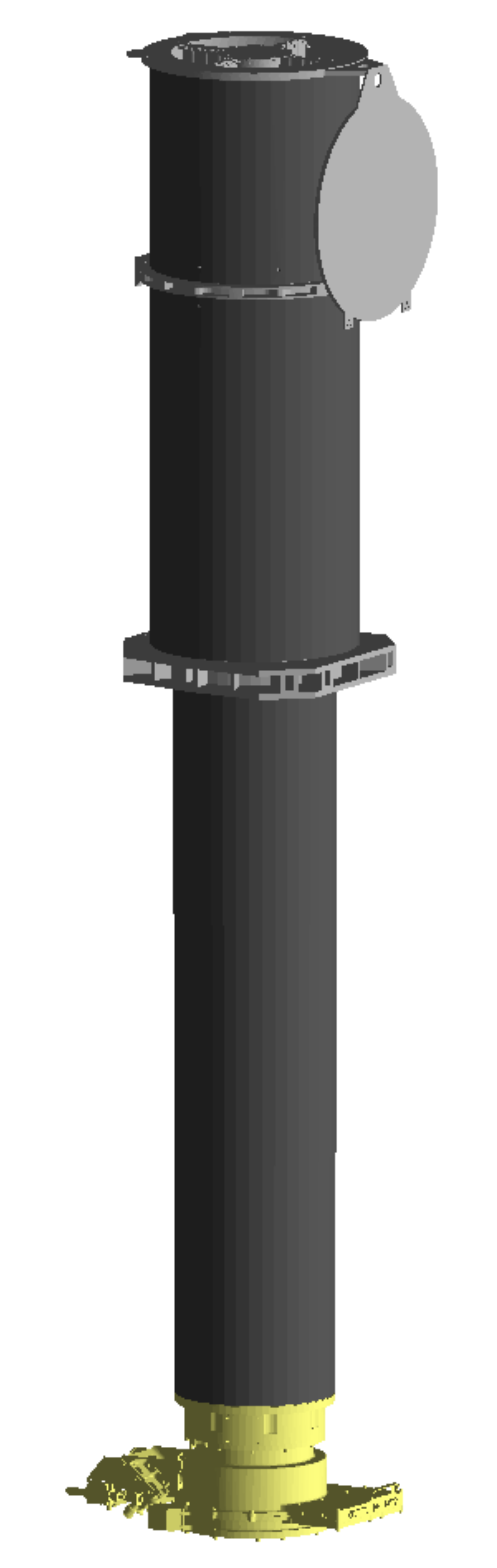}
    \caption{SXT}
    \label{fig:sxt}
\end{subfigure}
\parbox{0.4\textwidth}{
    \begin{subfigure}{\linewidth}
        \includegraphics[width=\linewidth]{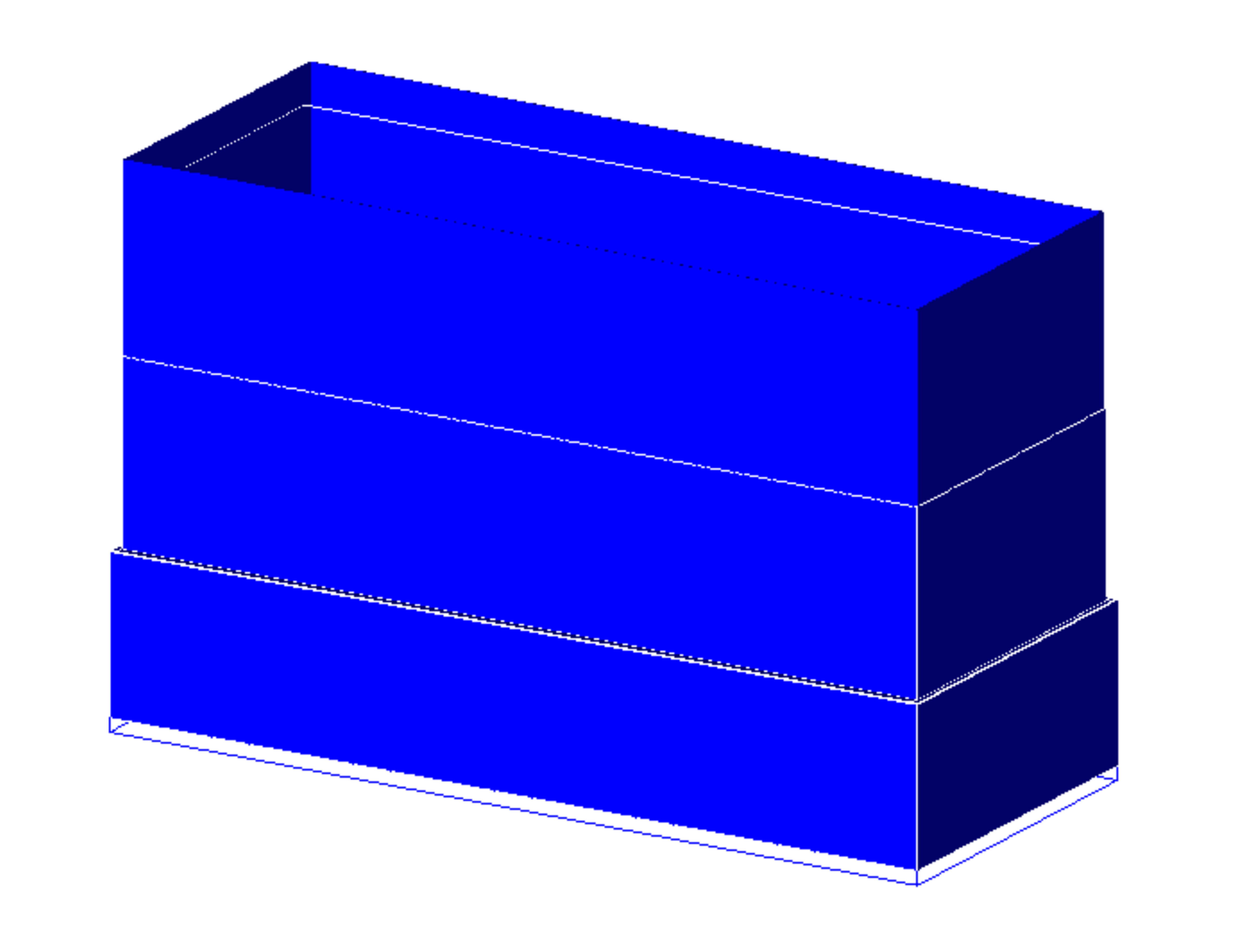}
        \caption{LAXPC}
        \label{fig:laxpc}
    \end{subfigure}\\
    \begin{subfigure}{\linewidth}
        \includegraphics[width=\linewidth]{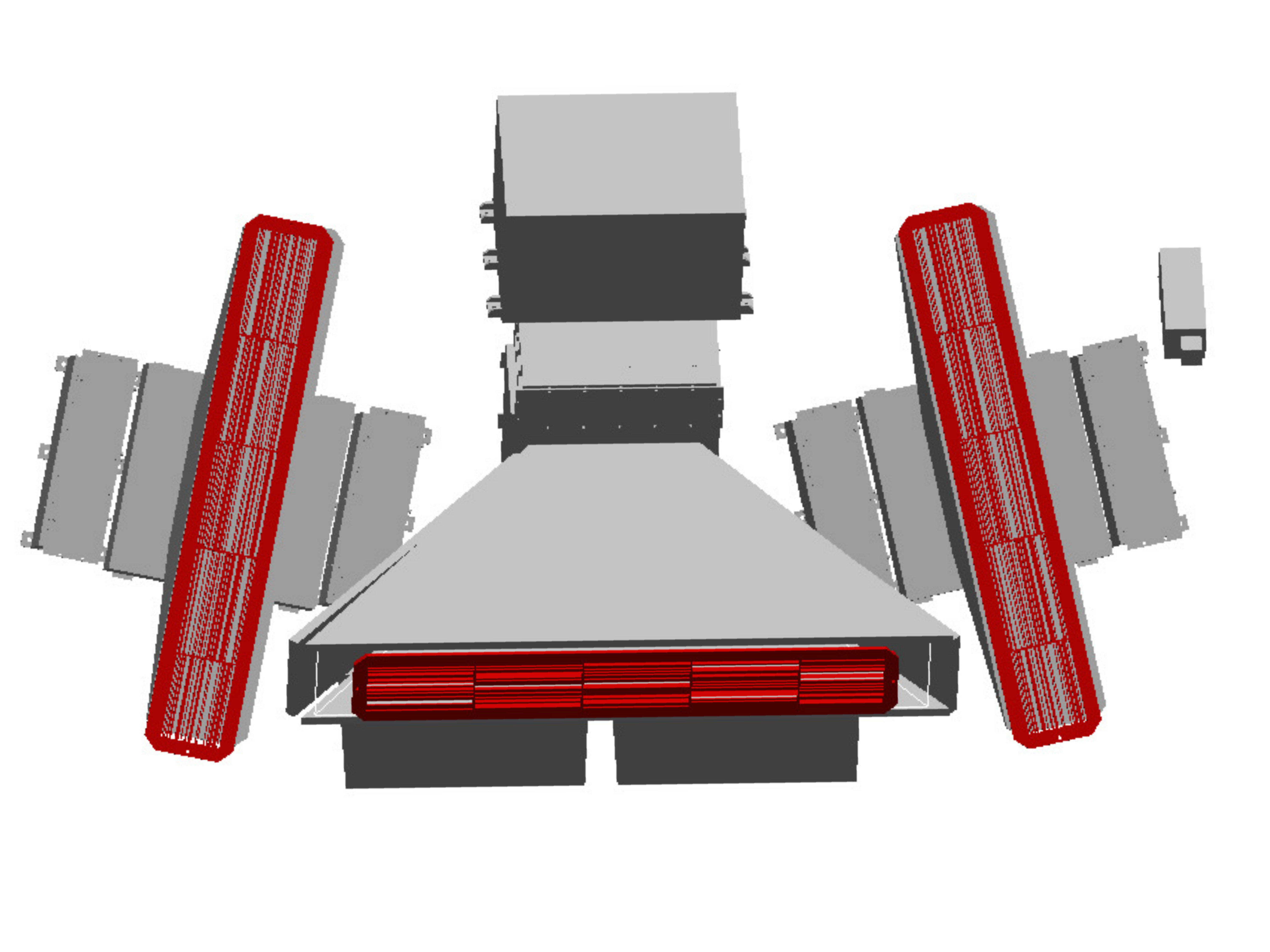}
        \caption{SSM}
        \label{fig:ssm}
    \end{subfigure}
}
\begin{subfigure}{0.25\textwidth}
    \centering
    \includegraphics[height=4in]{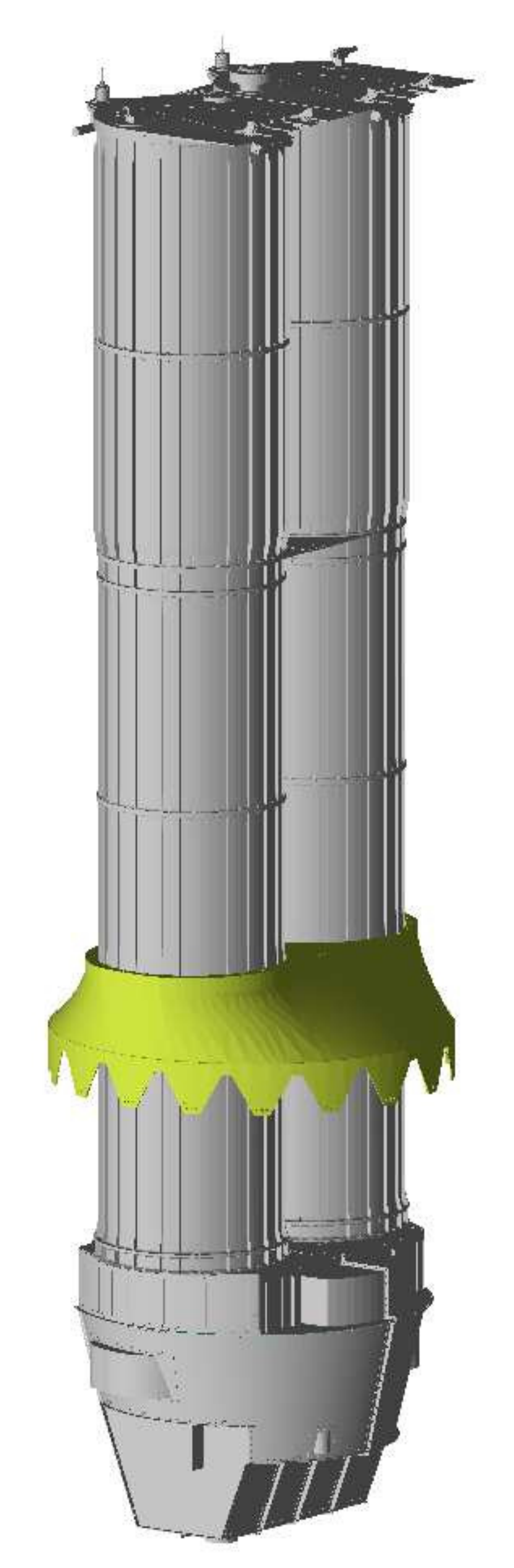}
    \caption{UVIT}
    \label{fig:uvit}
\end{subfigure}
\caption{\geant\ rendering of major \asat\ instruments. \label{fig:instruments}}
\end{figure*}

The mass model of LAXPC is taken from~\cite{laxpcMM} and the geometry is defined completely by \geant\ geometry classes (Figure~\ref{fig:laxpc}). A particularly complex system is the collimator, comprising of uniformly spaced parallel slats. However, the spacing between the slats ($\sim$ cm) is much smaller than the distance between the collimators and the CZT detectors ($>$ several tens of cm) which are the focus of our work. Hence, we can safely approximate the collimators of LAXPC as box of effective mass and composition to reduce the simulation time and for simplicity of incorporating the model in our code. All the other elements, except gas pump and electronic cards, are modelled accurately. The three units do not have the same gas: LAXPC 10 and 20 have 90\% Xenon and 10\% Methane, while LAXPC 30 has 84\% Xenon, 9.4\% Methane and 6.2\% Argon, all at 1520 torr pressure~\citep{astrosatHB}. This gives density of $1.07 \times 10^{-2}$~g~cm$^{-3}$ for LAXPC 10 and 20 and $1.22 \times 10^{-2}$~g~cm$^{-3}$ for LAXPC 30.


\subsection{Soft X-ray Telescope}\label{sec:sxt}
The geometry of SXT is completely imported as a CAD model, with some approximations for compositions of geometric parts (Figure~\ref{fig:sxt}). The focusing tube is made of Carbon Fibre Reinforced Polymer (CFRP), which we define as a new material with low density Carbon as the only element. The optics housing is pure Aluminium. The actual optics consist of concentric shells of Aluminium coated with Gold, but for simplicity we model them as a solid body composed of a Gold-Aluminium alloy with the same fractional abundances and total mass. Similarly in case of the camera assembly, the composition is kept as an alloy of Aluminium, Nickel and Gold with effective fraction and mass to incorporate the effect of Nickel and Gold coating as in real case.

\subsection{Ultraviolet Imaging Telescope}\label{sec:uvit}
Bulk of the UVIT geometry is imported from a CAD model (Figure~\ref{fig:uvit}). A notable exception is the highly complex geometry of the camera assembly, which we model with a \geant\ geometry class as two Aluminium cylinders with the same effective mass as the camera. The focusing tubes are made of two parts: Aluminium for parts above the top deck of the satellite, and Invar for the part between the top and bottom deck. The satellite adapter connecting the tubes and central cylinder between decks is of pure Titanium. The thermal blanket covering camera assembly is made of an Aluminium and Titanium alloy, where the Titanium fraction approximates the fasteners used in the assembly.

\subsection{Scanning Sky Monitor}\label{sec:ssm}
The SSM geometry is also imported from the CAD model (Figure~\ref{fig:ssm}). The body composition, along with the coded mask is Aluminium and the gas composition is 25\% Xenon and 75\% P10 (90\% Argon + 10\% Methane) at 800 torr pressure. The average density of the gas is $2.9\times10^{-3}$~g~cm$^{-3}$~\citep{astrosatHB}. SSM is mounted on a rotating platform, and it changes the orientation every ten minutes in routine \asat\ observations. However, we do not model this platform and keep the orientation of SSM fixed to the configuration at the time of the launch.

\subsection{Satellite Body}\label{sec:satbody}
Various parts of the satellite structure (satellite bus, support structures, electronics, etc.) significantly scatter incident photons. This effect is most prominent for GRBs shining onto CZTI from under the satellite ($-Z$ direction). This warrants a careful modelling of the satellite supporting structures as well as auxiliary electronics.

\begin{figure*}[ht!]
    \begin{subfigure}{0.5\textwidth}
        \includegraphics[width=\linewidth]{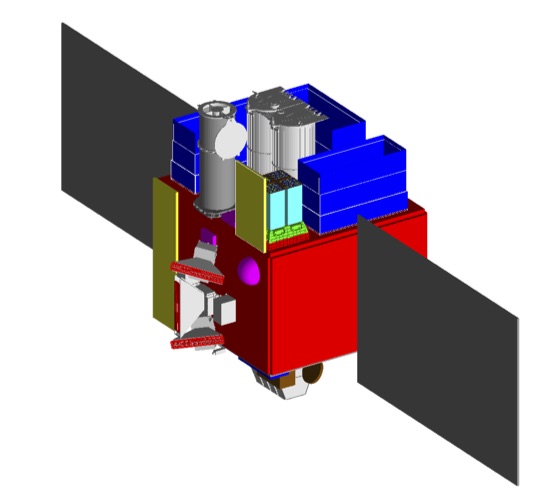}
        \caption{The entire satellite.}
        \label{fig:satfull}
    \end{subfigure}
    \begin{subfigure}{0.5\textwidth}
        \includegraphics[height=\linewidth]{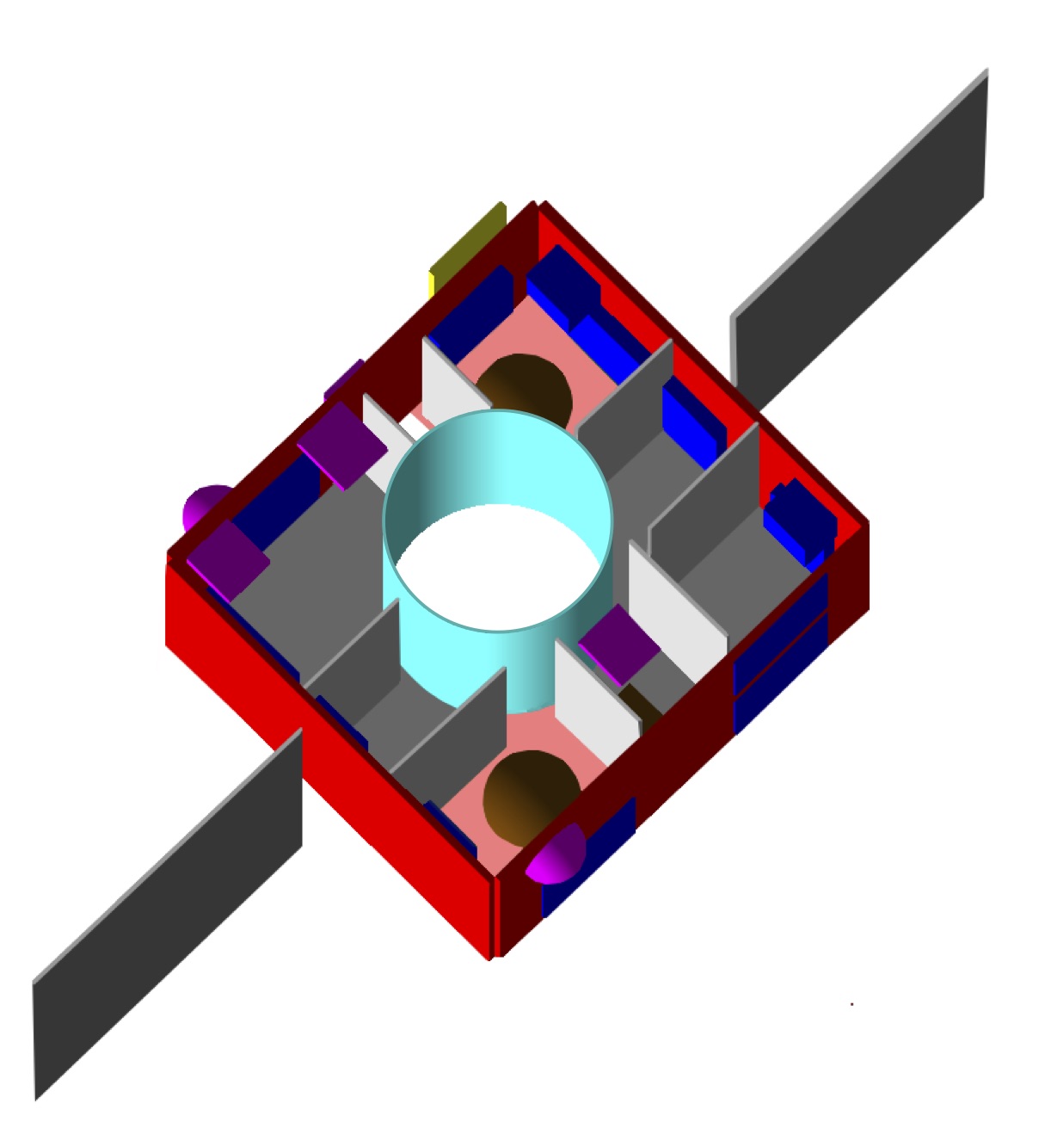}
        \caption{Inside view of the satellite body.}
        \label{fig:satinside}
    \end{subfigure}
\caption{\geant\ rendering of the satellite assembly. SSM and the solar panels are always assumed to be fixed at the orientation shown in this figure. \label{fig:body}}
\end{figure*}

This modelling is simplified by noting that if the satellite components are not too close to the CZTI detector plane, then small-scale structural details do not matter. For such components, only the  effective mass, composition and geometry is modelled accurately. This is done especially in case of electronic boxes on inside of the side plates of the satellite body. Instead of modelling each box, effective mass is split into separate boxes and they are placed uniformly on the side panels from inside. Other structures inside of the satellite such as fuel tanks, vertical support slats, and UVIT connector cylinder are modelled accurately with appropriate \geant\ geometry classes. Solar panels are approximated as a single sheet instead of two. The rotation with the orbit is not accounted while simulating and the orientation of panels is always vertical. Most of the satellite body is Aluminium honeycomb (lower density Aluminium). The electronic boxes are PCB plus Aluminium composite. Figure~\ref{fig:body} shows rendering of complete \asat\ geometry and top view into the inside of the satellite.

\section{Physics and Tracking}\label{sec:phys}

To define the particles and physics processes for the simulation, we employ user defined physics list derived from the \sw{G4VUserPhysicsList} class in \geant. As our interest is to obtain the response of CZTI for photon energies less than a few MeV, physics processes involving low  energy X-ray photons and electrons (secondary particles generated by photon interactions) are included in the physics list. In particular the following processes are used: \sw{G4Livermore\-PhotoElectricModel}, \sw{G4LivermorePolarized\-ComptonModel}, \sw{G4Livermore\-PolarizedRayleighModel}, \sw{G4Livermore\-IonisationModel}, \sw{G4Livermore\-BremsstrahlungModel}, and \sw{G4eMultiple\-Scattering}. Note that for Compton and Rayleigh scattering, models for polarised photons are added, as the mass model is also used to study the polarization characteristics of off-axis sources like GRBs~\citep{2019ApJ...884..123C}.

A \geant\ simulation ``Run'' includes generation of multiple seed photons and tracking their interaction with the volumes to obtain the required information. In \asat\ mass model simulations, CZT and veto CsI detectors are active volumes, for which the details of interaction and energy depositions are recorded for further analysis. The interactions of a given seed photon and the secondaries produced are part of an event of the Run. Energy depositions and interaction positions in the CZT and CsI detector volumes for each step of an event are accumulated. Here, the processing diverges for two different modes that we use in simulations.

The first mode is the ``response mode'', where our interest is to calculate the effective response of the satellite for incident photons, broadly equivalent to a combination of the ancillary response file (ARF) and the redistribution matrix file (RMF), but not including the detector effects like energy resolution.
In this mode, at the end of an event, for interactions in the CZT detectors, the pixel numbers are computed from the recorded interaction positions and total energy deposited in each pixel is calculated by adding up the energy depositions from all the interactions within a pixel. We only store events where the total energy is deposited in one pixel (single event) and discard the events where
the total energy is shared in multiple pixels (multiple events). The simulation output is saved as the total number of single events as a function of energy for each CZT pixel. For veto interactions, the total energy deposited is calculated and the net spectrum (number of events as a function of energy) is saved for each of the four veto detectors.

The second mode is a ``polarisation mode'', where it is important to process the photon interactions in further detail to identify Compton scattering double-pixel events for polarisation analysis. In this mode, instead of recording just the total energy deposit in a pixel, details of all the interactions (like xyz positions, energy deposition in each interaction, type of interaction etc.) for each incident photon (and its secondaries) within the detector volume are written out to a file. This file is then post-processed outside \geant ~for polarization analysis.


\section{Validation}\label{sec:validation}
\begin{figure*}[th]
\includegraphics[width=\textwidth]{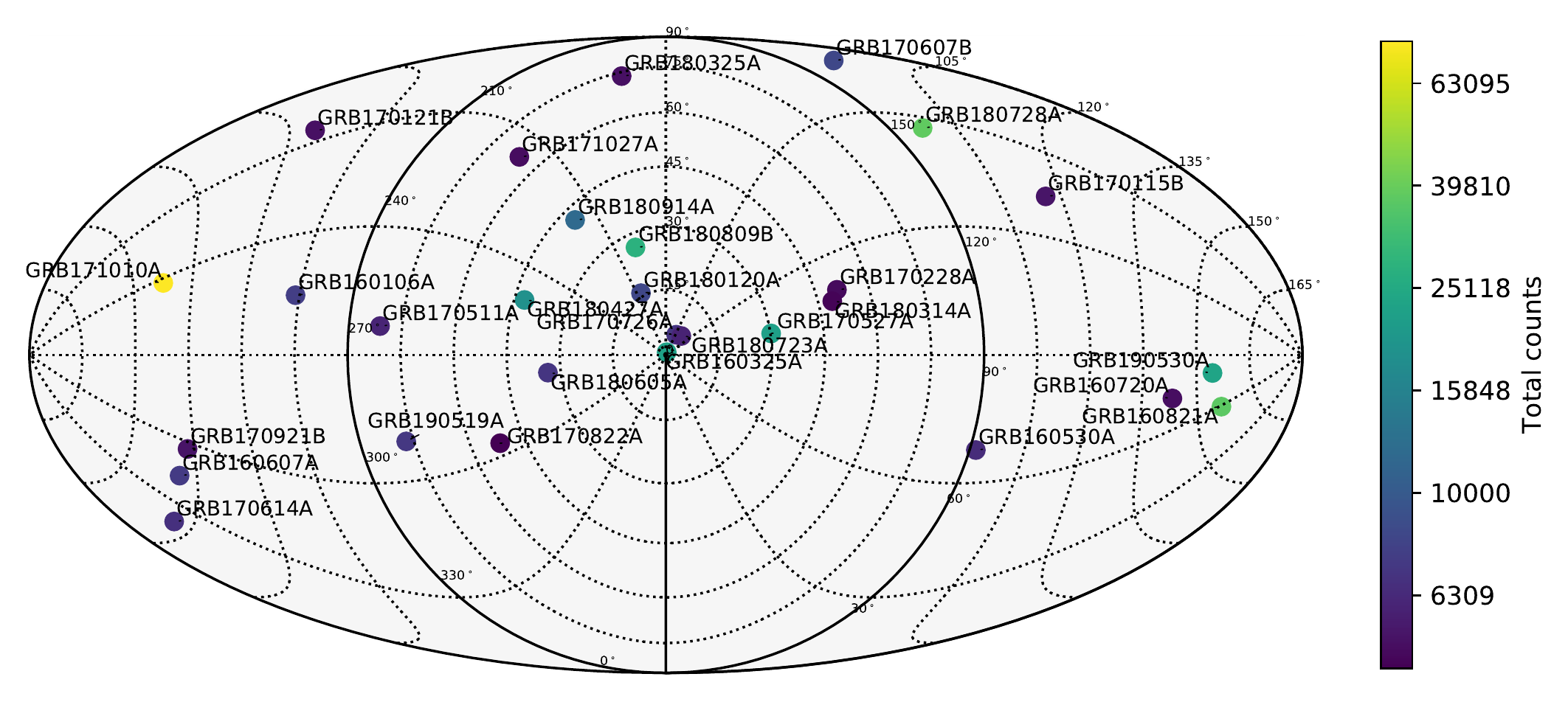}
\caption{GRB directions in CZTI reference frame. The mollweide projection is centred on the CZTI bore sight, such that $\theta = 0\degr, \phi = 0\degr$ ($Z$ axis) is at the centre of the figure. The concentric dotted circles show constant $\theta$ in steps of 15\degr, while the dotted lines originating from the centre are lines of constant $\phi$, in steps of 30\degr. The $X$ axis ($\phi=0$) points towards the bottom, and the $Y$ axis points towards the right. Colours denote the total number of counts detected in CZTI. \label{fig:grbdir}}
\end{figure*}

Validation of the mass model is crucial to understand its strengths and limitations. Like all satellites, the instruments of \asat\ were built separately and integrated together only at the final stage, leaving no opportunity for ground calibration with off-axis radioactive sources. Instead, we rely on astrophysical sources for validating the mass model. Since a large number of sources spread over the sky are contributing to the DPH and spectra, we cannot decompose the data into contributions from individual sources for such analysis. Instead, we rely on GRBs for the validation: we can estimate the DPH and spectrum of the GRB by subtracting the background contribution from all other sources as measured from time intervals just before and after the GRB. We then simulate the same GRBs by using incident spectra and fluxes from literature, and compare the simulation outputs with observations.

We note an important caveat here: GRB photons are scattered by the Earth's atmosphere, and some of these scattered photons reach the satellite, and this ``albedo'' flux can be as high has 30\% of the incident flux~\citep{Palit2021}. The observed magnitude of the effect depends on the direction-dependent sensitivity of the satellite, the spectrum of the GRB, and relative positions of the source, Earth, and the satellite. In extreme cases where the source is in a low sensitivity part of the sky while the Earth is in a high-sensitivity direction, the detected albedo component may even be higher than the direct GRB flux. The joint modelling of the Earth albedo and the satellite response will be taken up in a future work, here we simply note that the albedo effect may be responsible for some discrepancy in results.

\subsection{GRB selection}
The primary criterion for selecting GRBs for validation was having a large number of photons  ($>$5000) detected by CZTI\footnote{Note: This criteria forces our sample to consist of long GRBs only. The shortest duration GRB in the sample is $\sim30$~s and the longest one is $\sim200$~s.}. 
We took GRB positions reported by other missions and calculated the coordinates of each GRB in the CZTI coordinate system (Figure~\ref{fig:czti}). The angle $\theta$ is defined as the angle from the $Z$ axis, while $\phi$ is the azimuthal angle measured from the $X$ axis in the usual right-handed sense. We selected a subset of GRBs that were well-spread out in this $\theta$-$\phi$ coordinate system (Figure~\ref{fig:grbdir}, Table~\ref{tab:grbinfo}).

\begin{sidewaystable*}
\scriptsize
\caption{Spectral properties of all GRBs used in the validation analysis.\label{tab:grbinfo}}
\begin{tabular}{|c|cc|cccc|cccc|c|}
\hline
& \multicolumn{2}{|c|}{CZTI} & \multicolumn{8}{|c|}{Band Parameters} & \\
\cline{2-11}
GRB Name & \multicolumn{2}{c|}{Co-ordinates (deg)} & \multicolumn{4}{c|}{\fermi\ GBM} & \multicolumn{4}{c|}{\kw} & GCN reference\\
 & $\theta$ & $\phi$ &  $\alpha$ & $\beta$ & E$_{p}^\dag$ & N$^\ddag$ & $\alpha$ & $\beta$ & E$_{p}^\dag$ & N$^\ddag$ & (in addition to \fermi\ burst catalog) \\[4pt]
\hline
GRB160325A & 0.65 & 159.48 & $-0.74\pm0.04$ & $-2.32\pm0.15$ & $239.66\pm13.64$ & $0.01752$ & $-0.93^{0.15}_{-0.14}$ & $-2.45^{0.01}_{0.01}$ & $214^{31}_{-23}$ & $0.00774$ & \citet{GRB160325A_a}$^\star$\\[2pt]
GRB170726A & 5.46 & 147.88  & $-1.19\pm0.06 $ & $-1.81\pm0.12$ & $325.23\pm77.83$ & $0.00477$ & $-$ & $-$ & $-$ & $-$ & \\[2pt]
GRB180120A & 15.88 & 206.27  & $-1.09\pm0.01$ & $-2.63\pm0.08$ & $125.88\pm2.46$ & $0.06407$ & $-$ & $-$ & $-$ & $-$ & \\[2pt]
GRB180809B & 26.46 & 198.83  & $-$ & $-$ & $-$ & $-$ & $-0.87^{0.12}_{-0.1}$ & $-2.44^{0.19}_{-0.24}$ & $275^{25}_{-30}$ & $0.02210$ & \citet{GRB180809B_a}$^\star$,~\citet{GRB180809B_loc}$^\P$\\[2pt]
GRB170527A & 30.19 & 99.79  & $-1.16\pm0.00$ & $-5.40\pm8.87$ & $1094.61\pm46.52$ & $0.01415$ & $-0.71^{+0.09}_{-0.09}$ & $-2.5^?_?$ & $571^{55}_{-47}$ & $0.03144$ & \citet{GRB170527A_a}$^\star$,~\citet{GRB170527A_loc}$^\P$ \\[2pt]
GRB180605A & 33.9 & 277.37  & $-0.66\pm0.04$ & $-3.13\pm66.78$ & $639.83\pm49.97$ & $0.00959$ & $-$ & $-$ & $-$ & $-$ &  \\[2pt]
GRB180914A & 40.68 & 215.76 & $-$ & $-$ & $-$ & $-$ & $-0.68^{0.13}_{-0.12}$ & $-2.43$ & $301^{30}_{-25}$ & $0.11279$ & \citet{GRB180914A_a}$^\star$,~\citet{GRB180914A_loc}$^\P$\\[2pt]
GRB180427A & 42.17 & 250.99  & $-0.51\pm0.04$ & $-2.92\pm0.11$ & $110.07\pm2.65$ & $0.12519$ & $-0.68^{0.13}_{-0.12}$ & $-3.1^{0.17}_{-0.24}$ & $116^{4}_{-4}$ & $0.13346$ & \citet{GRB180427A_a}$^\star$,~\citet{GRB180427A_loc}$^\P$\\[2pt]
GRB180314A & 48.93 & 106.48  & $-0.40\pm0.07$ & $-3.36\pm1.47$ & $102.95\pm4.48$ & $0.07338$ & $-0.61^{0.17}_{-0.16}$ & $< -3.08$ & $111^{6}_{-6}$ & $0.03988$ & \citet{GRB180314A_a}$^\star$\\[2pt]
GRB170228A & 51.06 & 109.42  & $-0.79\pm0.04$ & $-4.67\pm9.50$ & $747.34\pm63.21$ & $0.00637$ & $-1.03^{+0.22}_{-0.18}$ & $< -1.98$ & $1017^{657}_{-317}$ & $0.00490$ & \citet{GRB170228A_a}$^\star$\\[2pt]
GRB190117A & 51.27 & 178.51  & $-$ & $-$ & $-$ & $-$ & $-1.19^{0.07}_{-0.08}$ & $-2.67^{0.27}_{-0.6}$ & $190^{18}_{-17}$ & $0.03198$ & \citet{GRB190117A_a}$^\star$,~\citet{GRB190117A_loc}$^\P$\\[2pt]
GRB170822A & 51.86 & 296.33 & $-$ & $-$ & $-$ & $-$ & $-1.2^{0.48}_{-0.38}$ & $< -2.4$ & $70^{12}_{-12}$ & $0.00804$ & \citet{GRB170822A_a}$^\star$,~\citet{GRB170822A_loc}$^\P$\\[2pt]

\hline
GRB171027A & 66.01 & 216.13 & $-$ & $-$ & $-$ & $-$ & $-1.21^{0.2}_{-0.35}$ & $-4^\S$ & $176^{47}_{-35}$ & $-$ & \citet{GRB171027A_a}$^\star$,~\citet{GRB171027A_loc}$^\P$\\[2pt]
GRB180325A & 73.53 & 188.27  & $-$ & $-$ & $-$ & $-$ & $-0.5^{0.21}_{-0.19}$ & $-2.65^{0.31}_{-1.06}$ & $306^{50}_{-39}$ & $0.04754$ & \citet{GRB180325A_a}$^\star$,~\citet{GRB180325A_loc}$^\P$\\[2pt]
GRB190519A & 77.14 & 290.53  & $-1.04\pm0.03$ & $-3.1\pm0.30$ & $108.00\pm3.00$ & $0.03341$ & $-1.05^{0.1}_{-0.09}$ & $<-3.05$ & $125^{7}_{-6}$ & $0.10023$ & \citet{GRB190519A_a}$^\star$, \citet{GRB190519A_b}$^\blacklozenge$\\[2pt]
GRB170511A & 81.22 & 263.31  & $-1.04\pm0.03$ & $-2.48\pm0.10$ & $86.87\pm3.93$ & $0.03595$ & $-$ & $-$ & $-$ & $-$ & \\[2pt]
GRB160530A & 91.72 & 67.94  & $-0.68\pm0.01$ & $-2.50\pm0.04$ & $181.77\pm2.77$ & $0.11191$ & $-0.93^{0.03}_{-0.03}$ & $< -3.5$ & $638^{36}_{-33}$ & $0.07075$ & \citet{GRB160530A_a}$^\star$,~\citet{GRB160530A_loc}$^\P$\\[2pt]
GRB170607B & 97.25 & 169.64 & $-0.75\pm0.02$ & $-2.71\pm0.23$ & $571.49\pm21.48$ & $0.02355$ & $-$ & $-$ & $-$ & $-$ & \\[2pt]
GRB180728A & 97.66 & 146.24  & $-1.54\pm0.01$ & $-2.46\pm0.02$ & $79.20\pm1.40$ & $0.28219$ & $-1.48^{0.12}_{-0.1}$ & $-2.57^{0.12}_{-0.18}$ & $97^{7}_{-6}$ & $0.08483$ & \citet{GRB180728A_a}$^\star$, \citet{GRB180728A_b}$^\blacklozenge$\\[2pt]
GRB160106A & 106.12 & 255.69 & $-0.62\pm0.04$ & $-2.14\pm0.07$ & $307.221\pm16.55$ & $0.02285$ & $-$ & $-$ & $-$ & $-$ & \\[2pt]
GRB170121B & 115.98 & 200.58  & $-1.03\pm0.02$ & $-2.68\pm1.61$ & $413.86\pm35.41$ & $0.00984$ & $-$ & $-$ & $-$ & $-$ & \citet{GRB170121B_loc}$^\P$\\[2pt]
GRB170115B & 116.27 & 132.60 &  $-0.88\pm0.01$ & $-2.47\pm0.36$ & $1563.00\pm106.00$ & $0.01406$ & $-0.86^{0.06}_{-0.06}$ & $-2.49^{0.31}_{-0.9}$ & $1114^{180}_{-156}$ & $0.05074$ & \citet{GRB170115B_a}$^\star$ \\[4pt]
\hline
GRB170921B & 136.68 & 302.73  & $-1.28\pm0.25$ & $-2.41\pm0.07$ & $88.39\pm3.53$ & $0.05564$ & $-1.42^{0.05}_{-0.05}$ & $< -3.3$ & $138^{7}_{-6}$ & $0.07566$ & \citet{GRB170921B_a} \\[2pt]
GRB170614A & 137.69 & 340.67  & $-1.12\pm0.04 $ & $-7.17\pm309.34$ & $348.96\pm28.35$ & $0.01749$ & $-$ & $-$ & $-$ & $-$ & \\[2pt]
GRB160607A & 138.85 & 315.77  & $-$ & $-$ & $-$ & $-$ & $-0.68^{1.47}_{-0.43}$ & $-2.51^{0.26}_{-0.35}$ & $176^{25}_{-42}$ & $0.02180$ & \citet{GRB160607A_a}$^\star$,~\citet{GRB160607_loc}$^\P$\\[2pt]
GRB171010A & 142.51 & 242.02  & $-1.09\pm0.005$ & $-2.19\pm0.01$ & $137.66\pm1.42$ & $0.11804$ & $-1.09^{0.04}_{-0.04}$ & $-2.42^{0.06}_{-0.07}$ & $171^{7}_{-6}$ & $0.11693$ & \citet{GRB171010A_a}$^\star$ \\[2pt]
GRB160720A & 143.38 & 73.06  & $-0.92\pm0.17$ & $-1.95\pm0.28$ & $153.84\pm0.00$ & $0.00939$ & $-0.87^{0.04}_{-0.04}$ & $-2.87^{0.14}_{-0.18}$ & $227^{9}_{-8}$ & $0.01888$ & \citet{GRB160720_a}$^\star$\\[2pt]
GRB190530A & 154.50 & 80.31  & $-1.00\pm0.01$ & $-3.64\pm0.12$ & $900.00\pm10.00$ & $0.19478$ & $-1.03^{0.02}_{-0.02}$ & $-3.03^{-0.36}_{0.22}$ & $848^{44}_{-33}$ & $0.08921$ & \citet{GRB190530A_a}$^\star$, \citet{GRB190530A_b}$^\blacklozenge$\\[2pt]
GRB160821A & 156.18 & 59.27 & $-1.05\pm0.00$ & $-2.30\pm0.02$ & $940.62\pm16.02$ & $0.03263$ & $-1^{0.02}_{-0.02}$ & $-1.99^{0.04}_{-0.04}$ & $710^{47}_{-44}$ & $0.02021$ & \citet{GRB160821A_a}$^\star$\\[2pt]
\hline
\end{tabular}\\[3pt]
Notes:
\begin{enumerate}\itemsep3pt\parskip0pt
    \item Lack of value imply no detection by the respective instrument;\hspace{5pt}$\dag$ Peak energy in keV;\hspace{5pt} $\ddag$ Band function norm at 100 keV in ph/cm$^2$/s/keV.
    \item CZTI coordinates are computed by converting the GRB RA/Dec to the CZTI frame. The RA/Dec are taken from GBM burst catalogue~\footnote{\url{https://heasarc.gsfc.nasa.gov/W3Browse/fermi/fermigbrst.html}}~\citep{vonKienlin2020} except for GRB180809B, GRB170527A, GRB180914A, GRB180427A, GRB190117A, GRB170822A, GRB171027A, GRB180325A, GRB160530A, GRB170121B, GRB160607A where they are taken from the GCNs cited with $\P$.
    \item \fermi\ GBM spectral parameters are also taken from the GBM burst catalogue~\citep{vonKienlin2020} except for GRB190519A, GRB180728A and GRB190530A where the reference are taken from the GCNs marked with $\blacklozenge$.
    \item \kw\ spectral parameters are taken from the GCNs marked with $\star$.
    \item For GRB171027A the GCN uses cut-off power-law (CPL) to fit the spectrum, however since we only use Band function, we use $\beta$ value (marked by $\S$) which reasonably approximates the CPL.
\end{enumerate}

\end{sidewaystable*}

Finally, to validate our simulations against the observed data, we needed knowledge of the flux and spectral parameters of the bursts. We observed that incident photons with energies as high as a few MeV can get down-scattered to the 20--200~keV range of interest. However, the number of such photons in typical GRBs is very low. Based on convergence studies, we concluded that considering the incident spectrum up to 2~MeV was sufficient for our simulations. Hence, we preferred using GRBs where the spectra had been modelled to high energies, which led us to select GRBs detected by either \fermi\ or \kw\ missions. We use the Band model parameters~\citep{1993ApJ...413..281B} reported in the literature to simulate the source spectrum. The normalisation is calculated using the total fluence of the burst, circumventing the problem of variation of count rates over time. The final list of selected GRBs along with references for spectral parameters is given in the Table~\ref{tab:grbinfo}.

\subsection{Simulations}\label{sec:sim}
Instead of implementing the Band function in \geant, we opted to simulate monochromatic photon beams at various energies, and synthesise the final output by numerically integrating over the outputs of these simulations. An advantage of this method is the possibility of reusing the same simulation outputs for any other incident spectral model, by simply changing the weights in the final co-addition. We choose an energy grid based on the energy resolution of CZTI. The simulations start at 20~keV, with 5~keV steps till 500~keV. The step size increases to 10~keV in the 500~keV -- 1~MeV range, followed by 20~keV steps in the 1~MeV -- 2~MeV range. For each energy, we simulate 6.97 million incident photons shining on the entire satellite by using a circular source plane of 200~cm radius, generating a photon flux of 55.46~photons~cm$^{-2}$.

Both observed data and simulations show significant flux in Tantalum fluorescence lines in the 55--65~keV region ($K\alpha_1 = 57.5$~keV, $K\alpha_2 = 56.3$~keV, $K\beta = 65.2$~keV). These fluorescence lines give almost no information about the energy or direction of incident photons. Hence, we ignore the 50--70~keV band in our analysis. The satellite structure is highly absorbing at energies below 50~keV, limiting the utility of that energy range as well. Thanks to these cuts, all analyses discussed in this section are for the CZT data in the 70--200~keV band.

Veto detectors are non-imaging detectors. These detectors have poorer spectral resolution and underwent limited ground calibration. Hence, we do not discuss veto data in this paper.

\subsection{Comparing observations and simulations}\label{sec:compare}
Here, we show the comparison between our observed and simulated data for a few selected GRBs. Details for the remainder of the sample are given in \ref{sec:othergrbs}.

\begin{figure*}[hp]
\centering

\begin{subfigure}{\textwidth}
  \centering
  \includegraphics[width=\linewidth]{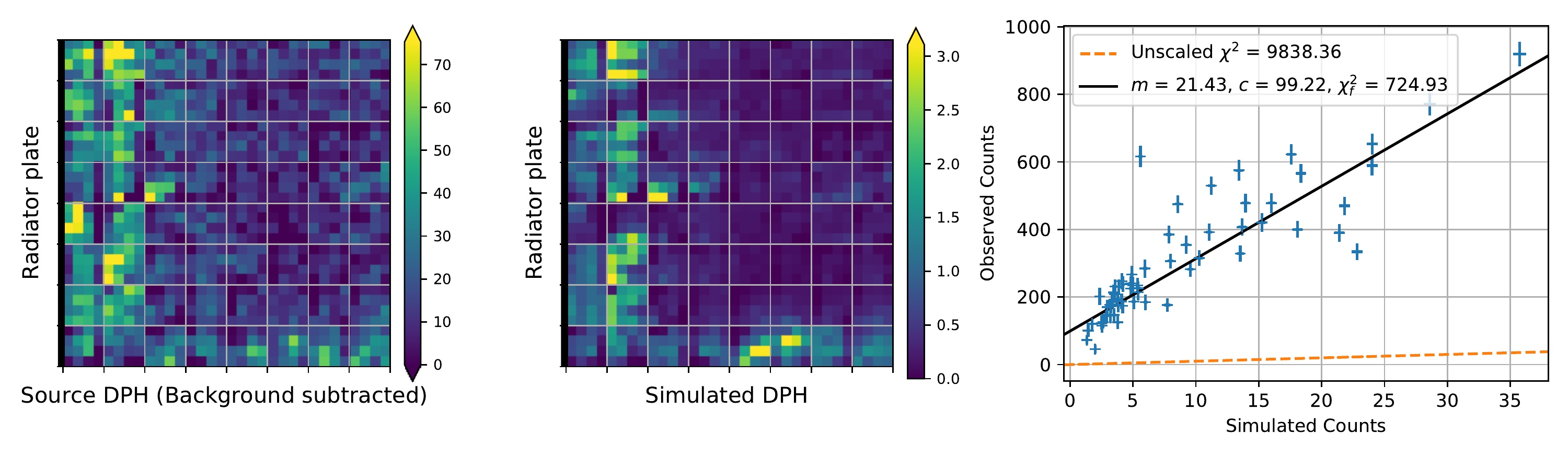}
  \caption{GRB180809B: $\theta=26.46\degr$, $\phi = 198.83\degr$}
  \label{fig:GRB180809B}
\end{subfigure}

\begin{subfigure}{\textwidth}
  \centering
  \includegraphics[width=\linewidth]{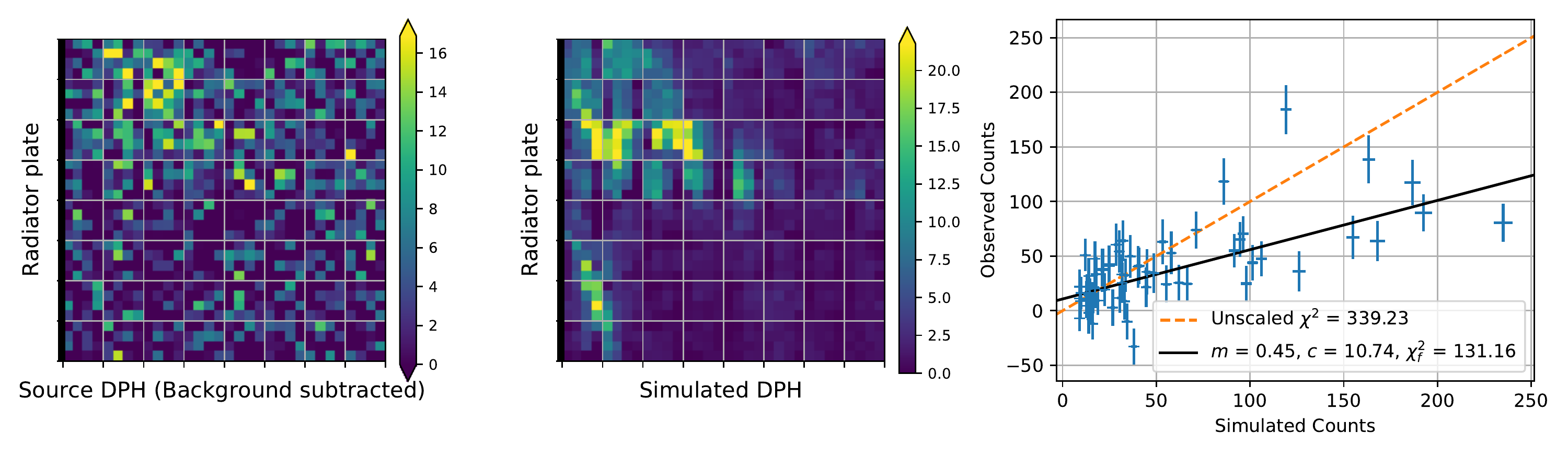}
  \caption{GRB180314A: $\theta=48.93\degr$, $\phi = 106.48\degr$}
  \label{fig:GRB180314A}
\end{subfigure}

\begin{subfigure}{\textwidth}
  \centering
  \includegraphics[width=\linewidth]{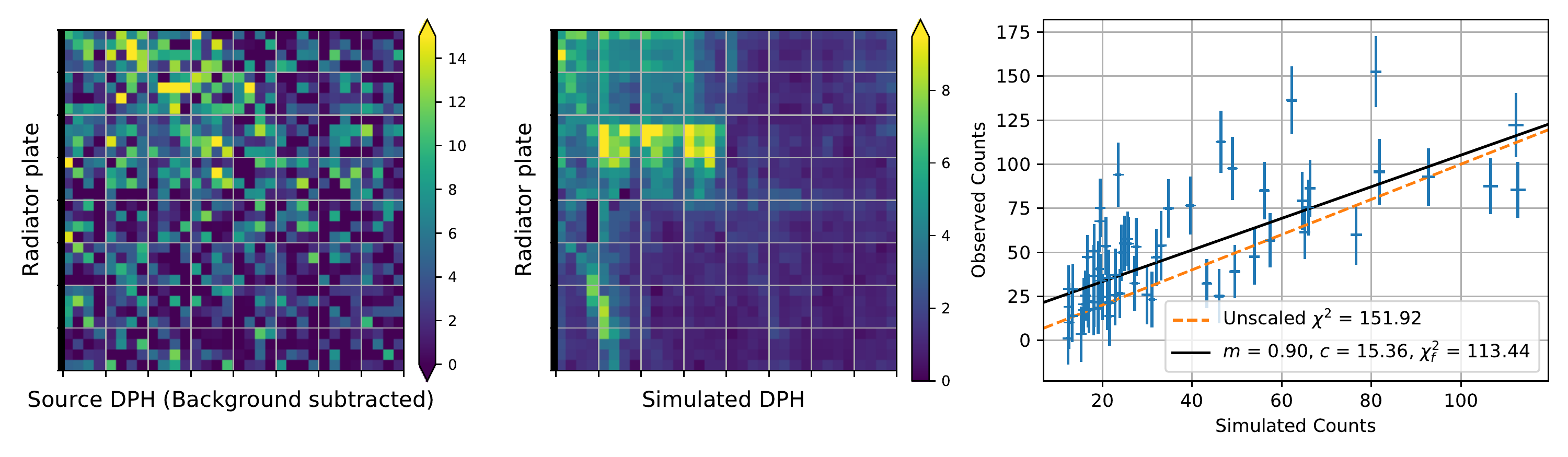}
  \caption{GRB170228A: $\theta=51.06\degr$, $\phi = 109.42\degr$}
  \label{fig:GRB170228A}
\end{subfigure}

\begin{subfigure}{\textwidth}
  \centering
  \includegraphics[width=\linewidth]{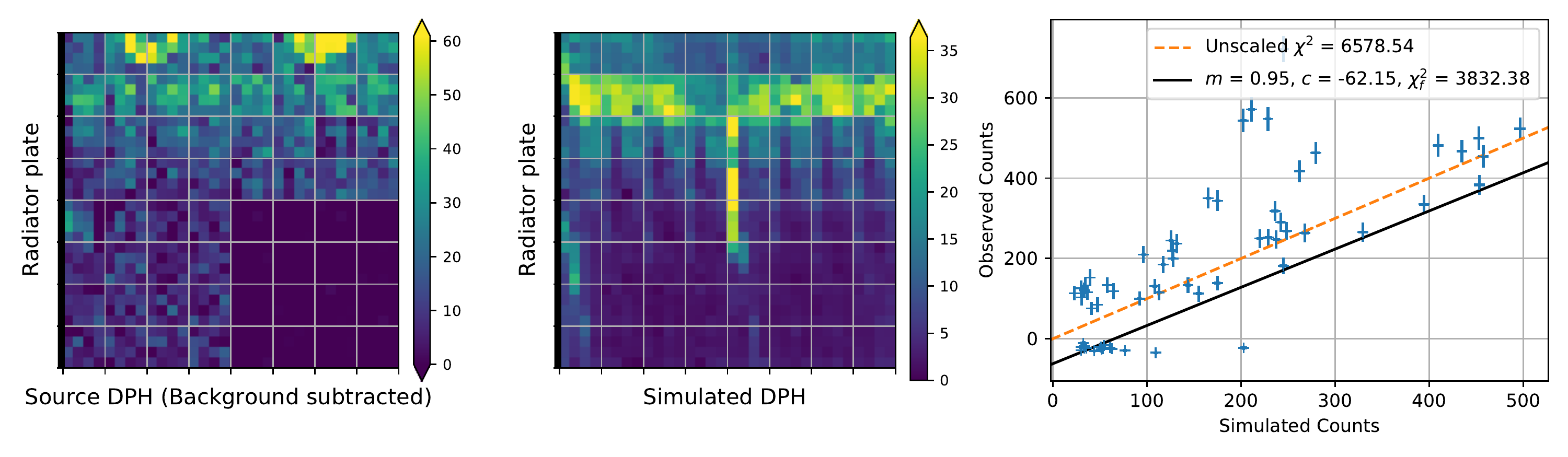}
  \caption{GRB170527A: $\theta=30.19\degr$, $\phi = 99.79\degr$}
  \label{fig:GRB170527A}
\end{subfigure}
\caption{Observed and simulated data for four GRBs above the detector plane. The left panels show observed, background-subtracted DPHs, while middle panels show simulated DPHs. These are binned in $4\times4$ pixel bins, and grey lines denote boundaries of detector modules. The DPHs are oriented such that the CZTI radiator plate is along the left side, and $\phi = 0\degr$ is along the $+X$ axis. In this orientation, quadrants are oriented in a clockwise manner with quadrant A at the top left adjacent to the radiator plate. For instance, in case of GRB170527A (\ref{fig:GRB170527A}), quadrant C data was unusable as explained in the text. Right panels show a scatter plot of module-wise observed versus expected counts, along with fits discussed in \S\ref{sec:quant}.} \label{fig:grbplots1}
\end{figure*}

\begin{figure*}[hp]
\centering

\begin{subfigure}{\textwidth}
  \centering
  \includegraphics[width=\linewidth]{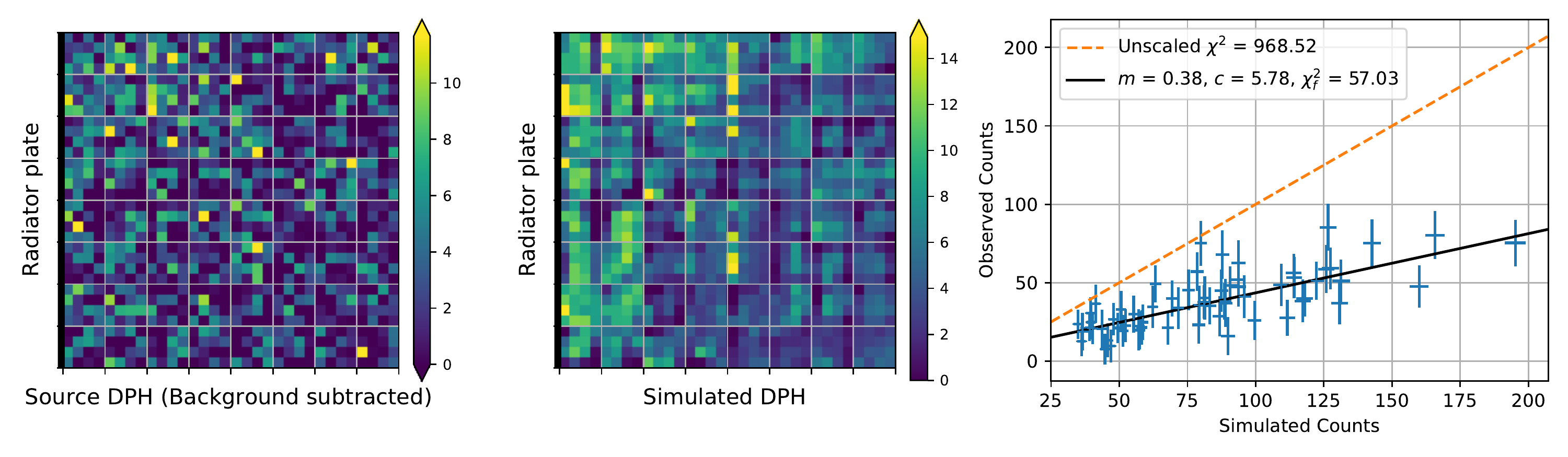}
  \caption{GRB180325A: $\theta=73.53\degr$, $\phi = 188.27\degr$}
  \label{fig:GRB180325A}
\end{subfigure}

\begin{subfigure}{\textwidth}
  \centering
  \includegraphics[width=\linewidth]{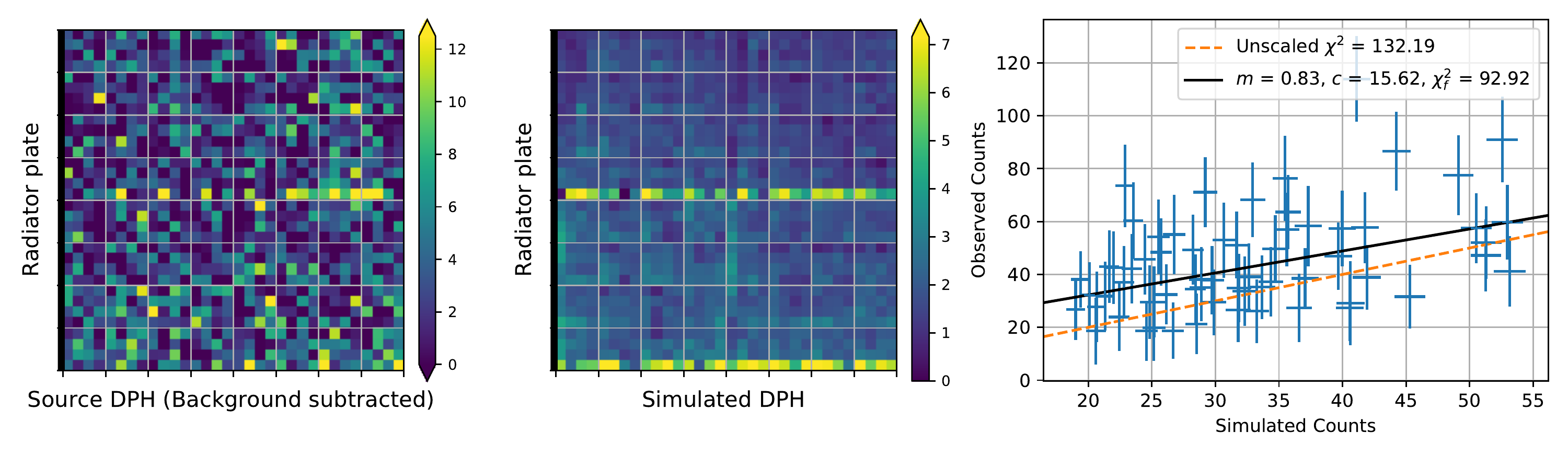}
  \caption{GRB170511A: $\theta=81.22\degr$, $\phi = 263.31\degr$}
  \label{fig:GRB170511A}
\end{subfigure}

\begin{subfigure}{\textwidth}
  \centering
  \includegraphics[width=\linewidth]{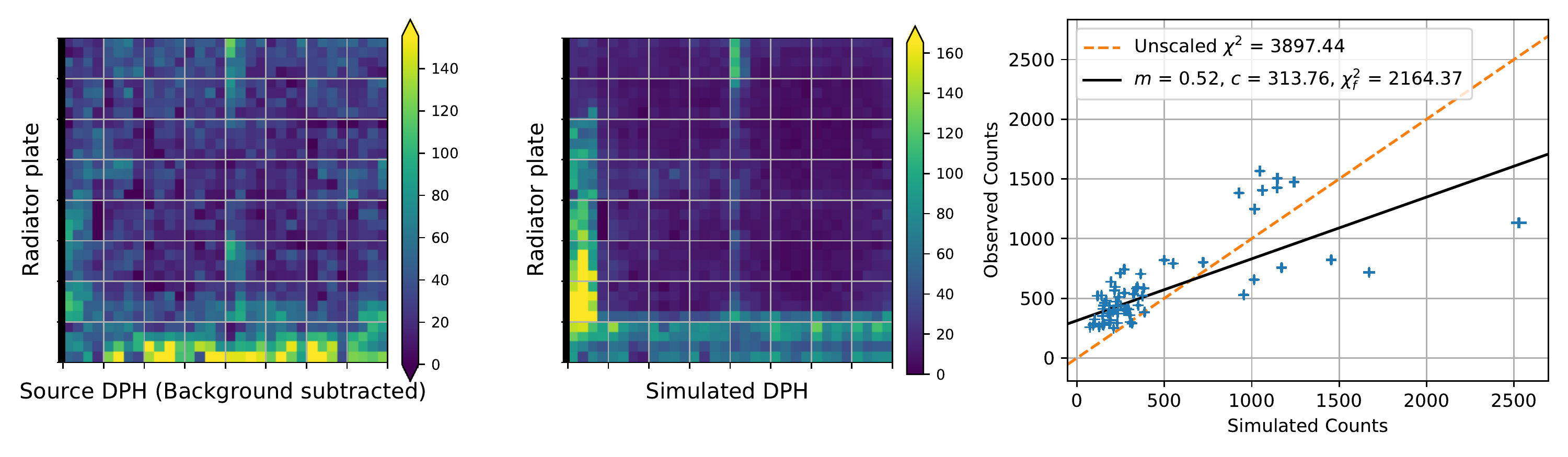}
  \caption{GRB171010A: $\theta=142.51\degr$, $\phi = 242.02\degr$}
  \label{fig:GRB171010A}
\end{subfigure}

\begin{subfigure}{\textwidth}
  \centering
  \includegraphics[width=\linewidth]{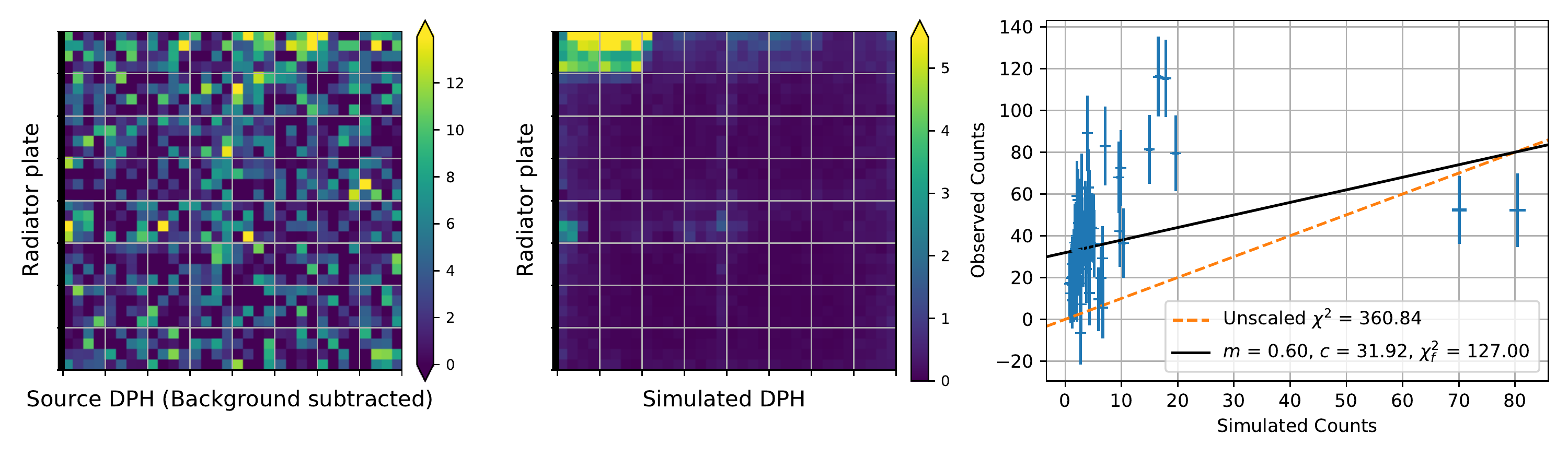}
  \caption{GRB160720A: $\theta=143.38\degr$, $\phi = 73.06\degr$}
  \label{fig:GRB160720A}
\end{subfigure}
\caption{Observed and simulated data for two GRBs incident at oblique angles, and two GRBs below the detector plane. Details are as in Figure~\ref{fig:grbplots1}.} \label{fig:grbplots2}
\end{figure*}

Figures~\ref{fig:grbplots1} and \ref{fig:grbplots2} show the comparison between observations and our mass model simulations for eight GRBs. The left and middle panels show the detector plane histograms (DPH) of the background-subtracted observed data and the mass model simulation respectively. We take a list of dead and disabled pixels from the CZTI pipeline, and set those pixels to zero in the simulations as well. The complete CZTI detector plane is a 256$\times$256 pixel grid. In these figures, we bin the DPH in 4$\times$4 pixel bins to visually suppress the inter-pixel Poisson variations. We plot the DPH as a single image, without rendering the inter-detector or inter-quadrant spacing seen in Figure~\ref{fig:czti2}. Grey lines denote the boundaries of individual detector modules. Some pixels in the source DPH can be negative as a result of background subtraction and Poisson noise. For visual clarity, the minimum value of the source DPH colour bar is set to zero, such that pixels with $\le0$ are rendered as purple. This problem does not affect the simulated DPHs, and we set the lower bound to zero for all of them. In all DPHs (observed and simulated), the upper limit of the colour bar is set to the 99th percentile of the DPH counts to prevent a few bright pixels from skewing the rendering. The right panels show comparisons of the observed and simulated counts in each detector module, and are discussed in detail in \S\ref{sec:quant}. We caution the readers about two effects here. First, there are unknown contributions from Earth albedo to these DPHs, but they are not discussed in this paper. Second, the background subtraction process leaves residual Poisson noise in the source DPH which is not present in the simulation. This is discussed in \ref{sec:vispoisson}.

Let us first discuss the qualitative comparisons of the observed and simulated DPHs in 
Figures~\ref{fig:grbplots1} and \ref{fig:grbplots2}. Figure~\ref{fig:grbplots1} shows the comparisons for GRBs incident from well above the detector plane ($\theta \lesssim 60\degr$). We see that the simulations reproduce observations quite well in most cases. We see excellent agreement in the brightness patterns in GRB~180809B (Figure~\ref{fig:GRB180809B}), including not just the vertical and horizontal bright bands, but also vertical dark lines within the bright band that are seen in the observed as well as in the simulated data. We note with satisfaction that for GRB~180314A and GRB~170228A which are separated by just 3\degr, both observed and simulated DPHs are similar (Figures~\ref{fig:GRB180314A}, \ref{fig:GRB170228A}). At the same time, some features stand out. For instance, Quadrant~C was saturated with noise at the time of GRB~170527A, rendering parts of the data unusable. This decreases the overall counts in that quadrant (Figure~\ref{fig:GRB170527A}), an effect which is not replicated in the simulation. The two bright spots at the top of the observed DPH seem to be related to the scattering from the alpha detector holders, but we could not replicate this feature in simulations.

Figure~\ref{fig:grbplots2} shows GRBs with incidence directions closer to or below the detector plane. GRBs incident at oblique angles show fewer features in the DPH. For instance, GRB~180325A (Figure~\ref{fig:GRB180325A}) simply shows a broad gradient from upper left to lower right direction in both observations and simulations. 
The observed horizontal row of bright pixels in GRB~170511A (Figure~\ref{fig:GRB170511A}) is the last row of pixels in quadrant B, and there is a large physical gap between these and the next row that has not been shown in the DPH. GRB photons coming at an angle of just 8.8\degr\ from the detector plane shine on the 5~mm vertical sides of these pixels, thus giving higher counts. Our simulations slightly overestimate this effect, and also create a bright row at the bottom of the DPH.

Observed and simulated DPHs for GRBs incident from below the detector plane ($\theta \gtrsim 120\degr$) show some peculiar features: in particular, some modules may end up being much brighter in the simulation than in observed data. Inspection of DPHs for GRB~171010A shows overall good agreement in the features: the presence of bright edges in the left and bottom of the figure, as well as a plus-shaped fainter feature running across the centre (Figure~\ref{fig:GRB171010A}). However, we find that the left edge is predicted to be significantly brighter in the simulations as compared to data. GRB~160720A shows a similar problem with two bright modules in the upper left corner of the simulated DPH (Figure~\ref{fig:GRB160720A}). If we ignore these two modules, the features match better: the fifth and sixth detector modules of the top row are a bit brighter, and there are fainter ``blobs'' just below the centre, and the middle of the left edge. The effects of excluding such detectors from the fit are discussed in \S\ref{sec:quant}.


\subsection{Comparison of count rates}\label{sec:quant}
Next, we undertake a quantitative analysis by comparing the observed and simulated counts in each detector (right panels in Figures~\ref{fig:grbplots1} and \ref{fig:grbplots2}). We define the $\chi^2$ metric for this comparison as:
\begin{equation}
\chi^2= \sum \frac{(N_\mathrm{sim} - N_\mathrm{obs})^2}{\sigma^2_\mathrm{sim} + \sigma^2_\mathrm{obs}}
\end{equation}
The terms in this equation warrant some explanation. The observed count rate ($N_\mathrm{obs}$) is calculated as the difference between the number of counts in the GRB time window minus the background counts rate. We assume that the count rate is Poisson, and calculate $\sigma^2_\mathrm{obs}$ by adding the errors in quadrature. Each simulation of a monochromatic beam with energy $E$ gives $N(E)$ detected photons. As discussed in \S\ref{sec:sim}, we calculate weights $w(E)$ based on the flux and spectrum of the GRB, and obtain the simulated counts by numerically integrating over $E$.
To calculate the total error, we assume that the uncertainty in $N(E)$ is $\sigma(E) = \sqrt{N(E)}$. The final uncertainty is calculated by applying the same weights adding these uncertainties in quadrature: $\sigma^2_\mathrm{sim} = \sum w(E) \sigma^2(E) = \sum w(E) N(E)$. We note that this uncertainty calculation is not exact, but it serves as a sufficient approximation for our purposes.

To quantify the visual comparisons in \S\ref{sec:compare}, we now focus on the plots showing observed versus simulated counts (right panels). The dashed orange lines are unity-slope lines, for cases when the observed counts are equal to the simulated counts. GRB~170228A (Figure~\ref{fig:GRB170228A}), GRB~170527A (Figure~\ref{fig:GRB170527A}), GRB~170511A (Figure~\ref{fig:GRB170511A}) are some of the GRBs where the points cluster around these lines, showing a good agreement between observations and simulations. Any offsets from such a line would indicate improper background subtraction.

In other cases like GRB~180809B and GRB~180325A (Figures~\ref{fig:GRB180809B}   and \ref{fig:GRB180325A}) we see a strong linear correlation between the points, but with a different slope. We fit a straight line to this plot, and evaluate a $\chi^2_f$ value for this fit. Since $\sigma_\mathrm{sim} \ll \sigma_\mathrm{obs}$, we ignore $\sigma_\mathrm{sim}$ in this calculation.
This fit gives us a scaling parameter (the slope $m$) and an intercept ($c$), which are listed for all simulated GRBs in Table~\ref{tab:grbfit}. The intercept is interpreted as a residual due to improper background subtraction, which may arise from the orbital variation in background~\citep{czti}. We note that the GRBs with the highest values of $c$ are those with large $\theta$ values. The intercept here is indicative of an overall poorer match between our simulations and observations.

\begin{figure}[hp]
\centering
\begin{subfigure}{\linewidth}
  \centering
  \includegraphics[width=0.95\linewidth]{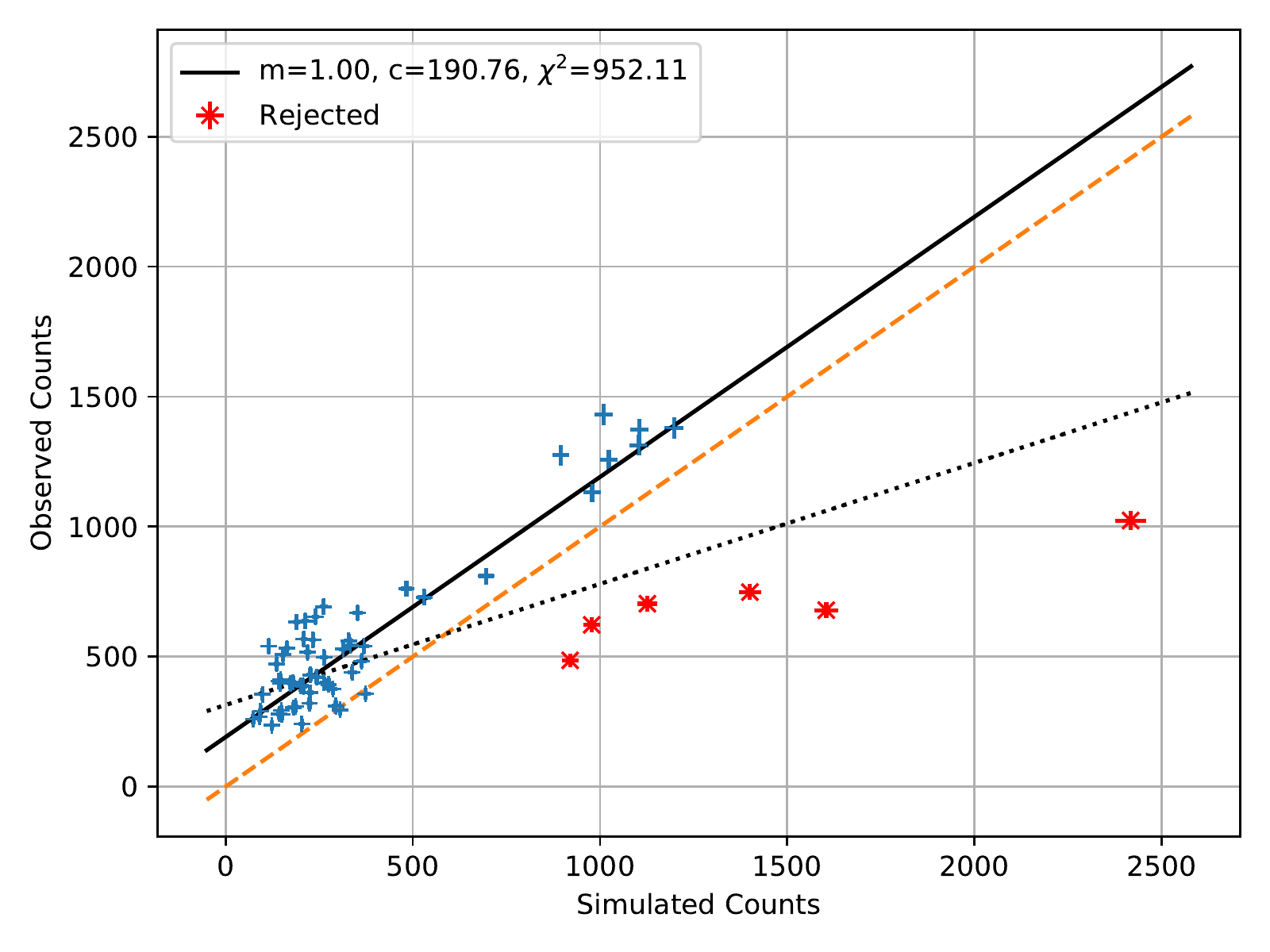}
  \caption{GRB~171010A: the exclusion of six bright modules along the left edge significantly improves the quality of the fit, decreasing the $\chi^2_f$ value from 2164 for 62 degrees of freedom to 952 for 56 degrees of freedom.}
  \label{fig:GRB171010_revised}
\end{subfigure}
\begin{subfigure}{\linewidth}
  \centering
  \includegraphics[width=0.95\linewidth]{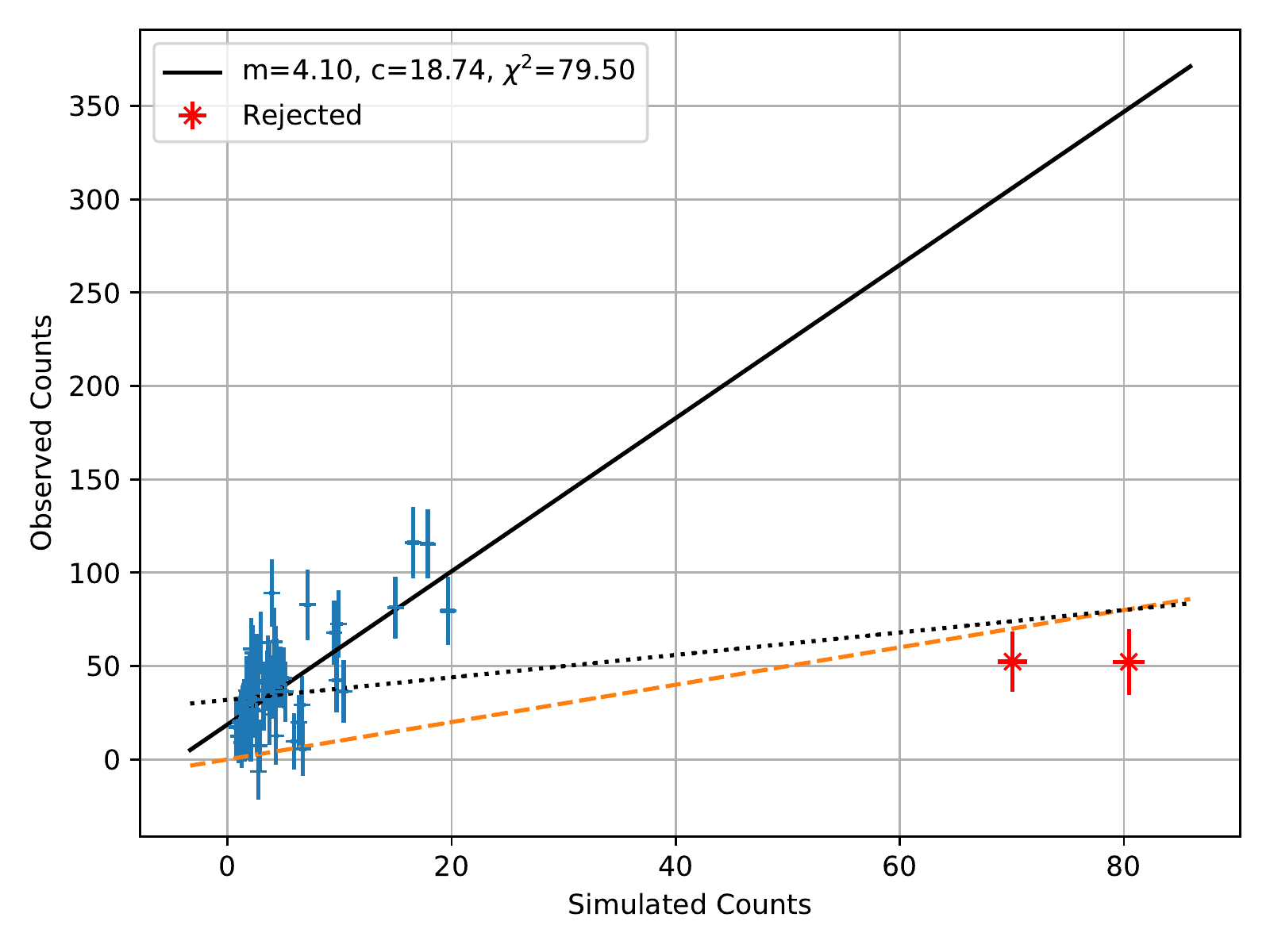}
  \caption{GRB~160720A: the exclusion two modules in the top left corner results in a decrease in the $\chi^2_f$ value from 127 for 62 degrees of freedom to 79 for 60 degrees of freedom.}
  \label{fig:GRB160720A_revised}
\end{subfigure}
  \caption{The original scaling relationships (dotted black line) for GRB~171010A and GRB~160720A are heavily skewed by simulated count rates of a few bright modules. Excluding these modules (red symbols) from the fit drastically improves the quality of the fit (solid black line). The dashed orange line denotes the ideal condition where observed counts are equal to predicted counts.}
\end{figure}

Some GRBs have a rather large $\chi^2_f$ value, indicative of a poor fit despite scaling (Table~\ref{tab:grbfit}). Two examples of this were discussed in \S\ref{sec:compare}: GRB~171010A and GRB~160720A (Figures~\ref{fig:GRB171010A},\ref{fig:GRB160720A}). The visual inspection of DPHs shows that counts in some modules have been over-predicted in both cases. For GRB~171010A if we ignore the six bright modules on the left side from the fit, we get a much better scaling relation ($m^\prime = 1.00$, $c^\prime = 191$) and the $\chi^2_f$ value decreases from 2164 for 62 degrees of freedom to 952 for 56 degrees of freedom (Figure~\ref{fig:GRB171010_revised}). A similar scenario is seen for GRB~160720A (Figure~\ref{fig:GRB160720A_revised}), where the exclusion of the upper two modules results in the $\chi^2_f$ decreasing from 127 for 62 degrees of freedom to 79 for 60 degrees of freedom. The scaling factor changes from $m = 0.60 \pm 0.16$ to $m^\prime = 4.1\pm0.5$.

We carefully examined the simulated data for such cases and found abnormalities in data quality. We infer that these abnormally bright patches, which strongly depend on incident direction, are a result of limitations of our mass model. Such artefacts could arise from missing components in the mass model, and can be considered as systematic errors. However, we need to characterise them and create objective criteria for excluding certain areas of the DPH for radiation incident from a given direction.

%
%



In other cases, we see that the slope of our best-fit is different from unity: for instance, $m = 0.45$ for GRB~180314A (Figure~\ref{fig:GRB180314A}), and $m = 0.38$ for GRB~180325A (Figure~\ref{fig:GRB180325A}). While the bright spots seen for some simulations indicate there although there are minor discrepancies in the mass model, the linear correlation demonstrates the validity of the placement of various components in the mass model and their relative opacities to the incoming radiation for these GRB directions.
While detailed spectral modelling required to measure CZTI flux is beyond the scope of this paper, we can use our scaling parameter $m$ as a proxy for ratio of fluence measured by CZTI to that of the reference spectral model. 
We note that no strong correlation is seen between the scaling parameter and the $\theta$ or $\phi$ coordinates of the GRB.

\begin{table*}
\centering
\caption{Best-fit scaling parameters and reduced $\chi^2$ values for detector-by-detector comparisons for all GRBs.\label{tab:grbfit}}
\begin{tabular}{|l|rr|c|cc|cc|}
\hline
GRB Name & $\theta$ & $\phi$ & Spectral Parameter Source & $m$ & $c$ & $\chi^{{2}}$ & $\chi^{{2}}_{{f}}$ \\
\hline
GRB160106A & 106.12 & 255.69 & \fermi & 0.48 & 6.46 & 767.87 & 151.07 \\
GRB160325A &   0.65 & 159.48 & \fermi & 0.16 & 70.55 & 15132.10 & 172.43 \\
GRB160530A &  91.72 &  67.94 & \fermi & 0.12 & 48.93 & 243.43 & 88.49 \\
GRB160607A & 138.85 & 315.77 & \kw & 1.32 & 30.81 & 421.51 & 105.95 \\
GRB160720A & 143.38 &  73.06 & \fermi & 0.24 & 31.88 & 372.18 & 126.78 \\
GRB160821A & 156.18 &  59.27 & \fermi & 1.22 & 222.51 & 6078.98 & 454.69 \\
GRB170115B & 116.27 & 132.60 & \fermi & 0.40 & 21.37 & 454.12 & 120.01 \\
GRB170121B & 115.98 & 200.58 & \fermi & 0.41 & 23.13 & 241.40 & 106.95 \\
GRB170228A &  51.06 & 109.42 & \fermi & 0.90 & 15.36 & 151.92 & 113.44 \\
GRB170511A &  81.22 & 263.31 & \fermi & 0.83 & 15.62 & 132.19 & 92.92 \\
GRB170527A$^*$ &  30.19 &  99.79 & \fermi & 0.87 & 82.5 & 1129.58 & 736.28 \\
GRB170607B &  97.25 & 169.64 & \fermi & 0.64 & 2.78 & 414.50 & 110.57 \\
GRB170614A & 137.69 & 340.67 & \fermi & 0.20 & 16.89 & 7147.74 & 311.75 \\
GRB170726A &   5.46 & 147.88 & \fermi & 0.30 & 47.69 & 274.68 & 72.85 \\
GRB170822A &  51.86 & 296.33 & \kw & 0.72 & 10.30 & 78.83 & 73.91 \\
GRB170921B & 136.68 & 302.73 & \fermi & 0.46 & 10.58 & 291.65 & 122.17 \\
GRB171010A & 142.51 & 242.02 & \fermi & 0.52 & 313.76 & 3897.44 & 2164.37 \\
GRB171027A &  66.01 & 216.13 & \kw & 1.16 & 14.12 & 103.07 & 61.32 \\
GRB180120A &  15.88 & 206.27 & \fermi & 0.20 & 21.36 & 3955.80 & 182.13 \\
GRB180314A &  48.93 & 106.48 & \fermi & 0.45 & 10.74 & 339.23 & 131.16 \\
GRB180325A &  73.53 & 188.27 & \kw & 0.38 & 5.78 & 968.52 & 57.03 \\
GRB180427A &  42.17 & 250.99 & \fermi & 0.75 & 20.72 & 726.79 & 645.48 \\
GRB180605A &  33.90 & 277.37 & \fermi & 0.41 & 20.67 & 485.38 & 129.84 \\
GRB180728A &  97.66 & 146.24 & \fermi & 0.19 & -4.91 & 8250.68 & 102.09 \\
GRB180809B &  26.46 & 198.83 & \kw & 21.43 & 99.22 & 9838.36 & 724.93 \\
GRB180914A &  40.68 & 215.76 & \kw & 0.10 & 42.54 & 39249.08 & 342.96 \\
GRB190117A &  51.27 & 178.51 & \kw & 0.54 & 23.68 & 519.76 & 145.84 \\
GRB190519A &  77.14 & 290.53 & \fermi & 0.81 & 8.56 & 132.49 & 128.02 \\
GRB190530A & 154.50 &  80.31 & \fermi & 0.13 & 115.62 & 6377.13 & 1104.98 \\
\hline
\end{tabular}
\tablenotes{$^*$ Quadrant C was extremely noisy at the time of GRB~170527A, and was excluded from the fit.}
\end{table*}

It is then likely that these slopes are indicative of the uncertainty in the source properties itself. These may be measurement uncertainties, or uncertainties in extrapolating from the energy range of the detecting instrument to the CZTI energy range. To illustrate this case, we consider the twelve GRBs from our sample where spectral parameters are known from both \fermi\ and \kw\ data. On comparing the incident flux in the 70-200~keV range using both spectral models (Table~\ref{tab:cbfr}), we find that while the inferred fluence values are in good agreement for about half the GRBs, in extreme cases they disagree by up to a factors of 9.3 (GRB~170921B) or even 55 (GRB~160720A). These discrepancies may arise from differences in the spectral parameters due to the different energy ranges of the instruments. The energy range and sensitivity differences even lead to different T$_{90}$ values for the two instruments, and thus some differences in the fluence values calculated by different instruments are unsurprising. Extending the same argument, even for a perfect CZTI Mass Model and simulation, we may expect disagreements of a similar magnitude when compared \fermi\ or \kw.
We see that while the flux estimates from \kw\ are often higher than those from \fermi\ spectral parameters, the range of values of $m$ is similar to that of the ratio of \kw\ and \fermi\ fluence values.

\begin{table*}
\caption{Comparing the 70--200~keV fluence values from \fermi\ and \kw\ parameters, in units of erg~cm$^{-2}$. We used \fermi\ spectral parameters for simulations of all these GRBs. The last two columns give the ratio of the the \fermi\ fluence to \kw\ fluence, and the multiplication factor required for matching \fermi-derived count rates to observed CZTI data.\label{tab:cbfr}}
\centering
\centering
\begin{tabular}{ccccc}
\hline
GRB Name & \fermi  & \kw & Ratio(KW/Fermi) & $m$\\
\hline
GRB160325A &	$6.11\times10^{-6}$ &	$3.81\times10^{-6}$ &	0.62 & 0.16\\
GRB160530A &	$3.43\times10^{-5}$ &	$4.43\times10^{-5}$ &	1.29 & 0.12\\
GRB160720A &    $8.17\times10^{-7}$ &   $4.54\times10^{-5}$ &   55.6 & 0.24\\
GRB160821A &	$1.04\times10^{-4}$ &	$4.31\times10^{-4}$ &	4.11 & 1.22\\
GRB170115B &	$1.17\times10^{-5}$ &	$4.77\times10^{-5}$ &	4.09 & 0.40\\
GRB170228A &	$4.28\times10^{-5}$ &	$1.83\times10^{-5}$ &	3.79 & 0.90\\
GRB170527A &    $1.80\times10^{-5}$ &   $2.96\times10^{-5}$ &   1.65 & 0.95\\
GRB170921B &	$2.02\times10^{-5}$ &	$1.88\times10^{-5}$ &	9.28 & 0.46\\
GRB171010A &	$2.06\times10^{-4}$ &	$1.76\times10^{-4}$ &	0.85 & 0.52\\
GRB180314A &    $7.53\times10^{-6}$ &   $5.95\times10^{-6}$ &   0.79 & 0.45\\
GRB180427A &	$2.07\times10^{-5}$ &	$2.12\times10^{-5}$ &	1.02 & 0.75\\
GRB180728A &	$1.61\times10^{-5}$ &	$1.54\times10^{-5}$ &	0.96 & 0.19\\
GRB190519A &	$1.17\times10^{-5}$ &	$1.21\times10^{-5}$ &	1.03 & 0.81\\
GRB190530A &	$6.99\times10^{-5}$ &	$7.00\times10^{-5}$ &	1.00 & 1.13\\
\hline
\end{tabular}
\end{table*}

\subsection{Spectral comparisons}
Besides the count rate comparisons as demonstrated in the previous section, we also attempted to validate the mass model by performing broad band spectroscopy for a number of GRBs using \fermi, {\em Swift} and \asat-CZTI data, where the spectral response file for CZTI is generated from the \geant\ simulations of full {\em Astrosat} mass model. Using an older version of this mass model with minor differences, a satisfactory agreement in spectral parameters between {\em Fermi} and CZTI was reported in \citet{2019ApJ...884..123C} for GRB 160821A. Paper II uses this latest version of the mass model to expand the work to eleven bright GRBs detected in the first year of CZTI operation and showcases capability of CZTI as a sub-MeV spectrometer, and thereby validating this \asat-CZTI mass model further.

\section{Discussion}\label{sec:conclusion}
The mass model presents a substantial development over the raytrace codes used in \citet{czti}: extending our understanding of CZTI sensitivity from $\sim$30\% of the sky to the entire visible sky. 
The all-sky median effective areas are 32.8~cm$^2$, 74.6~cm$^2$, and 68.2~cm$^2$ at 60~keV, 120~keV and 180~keV respectively. \citet{czti} considered only 30\% of the sky and calculated the median effective area at 180~keV to be 190~cm$^2$. Based on the more detailed physics of the mass model, this number decreases to 115~cm$^2$ for the same part of the sky. 
%
%

The conversion of effective area to sensitivity depends on the source spectrum, duration, as well as detector noise properties. \citet{sharma2020search} report that the typical minimum detectable count rate for CZTI is 284~counts~s$^{-1}$ for a 1~s burst, and 42~counts~s$^{-1}$ (total 420~counts) for a burst with a 10~s duration.  These count rates can be converted into direction-dependent sensitivity by assuming a source spectrum. As a proxy for an all-sky calculation, we evaluate the sensitivity in the directions of the 29 GRBs, where we have conducted mass model simulations.

For short GRBs, we consider a spectrum defined by a Band function with $\alpha = -0.5$, $\beta = -2.25$, and $E_\mathrm{peak} = 800$~keV \citep{2015MNRAS.448.3026W}. The light curve is assumed to comprise of a top-hat pulse with a width of 1~s. For such a burst, the CZTI fluence sensitivity ranges from $\sim8\times10^{-8}\mathrm{~erg~cm}^{-2}$ to $\sim2\times10^{-6}\mathrm{~erg~cm}^{-2}$, with a median value of $4\times10^{-7}\mathrm{~erg~cm}^{-2}$. 
For long GRBs, we consider a spectrum defined by a Band function with $\alpha = -1$, $\beta = -2.25$, and $E_\mathrm{peak} = 511$~keV \citep{2010MNRAS.406.1944W} 
and a 10~s top-hat pulse to get fluence sensitivities in the range from $1\times10^{-7}\mathrm{~erg~cm}^{-2}$ to $4\times10^{-6}\mathrm{~erg~cm}^{-2}$ with a median value of $7\times10^{-7}\mathrm{~erg~cm}^{-2}$. The range of these values is comparable to \fermi~GBM. Detailed analyses of GRBs detected by CZTI and quantification of the sensitivity will be addressed in a later work.

We discuss a few observations noted during our studies. For instance, we note that the DPH is not sensitive to the small changes in the input spectrum. Furthermore, Ta fluorescence lines dominate the observed spectrum in the 50 -- 70 keV energy range making it difficult to ascertain the source spectrum in this energy range. Finally, for very accurate analysis of GRBs coming from the SSM direction exact position the SSM should be taken into account as the orientation of SSM may affect the observed counts distribution. However, such GRBs would have a high $\theta$, and we have found that such cases show higher disagreements between observations and simulations, independent of $\phi$.

An older version of \asat\ CZTI mass model with minor differences has been successfully used to study the polarisation of prompt emission in several GRBs \citep{2016ApJ...833...86R,2019ApJ...884..123C,2019ApJ...874...70C}. \todo{Singhal et al., this issue} use the same older version to measure fluxes of bright sources by the earth occultation technique. Paper II demonstrates the validity of the mass model simulations for spectroscopic analyses, using this current mass model version.

Simulations of GRBs using the \asat\ CZTI mass model show satisfactory match with observed data. Here, we have demonstrated the good correspondence between observed and simulated count rates and DPHs. Our simulations with earlier versions of the mass model show that DPHs can show measurable differences over scales of $\sim 10\degr$. This can be leveraged for localising GRBs. To aid this work, we will calculate the mass model response over the entire sky using the vertices of a level 4 HTM grid, with a nominal spacing of 5.6\degr. Preliminary testing shows that by comparing simulated DPHs for various points on this grid with the observed DPH, we can localise  GRBs to $\lesssim 20\degr$ on the sky. Accurate localisation techniques will also have to correctly account for the effect of Earth albedo on the observed distribution of source photons on the detector plane. The same grid will also be utilised for future studies of counterparts to fast radio bursts and gravitational wave sources.



\section*{Acknowledgements}

CZT--Imager is built by a consortium of Institutes across India. The Tata Institute of Fundamental Research, Mumbai, led the effort with instrument design and development. Vikram Sarabhai Space Centre, Thiruvananthapuram provided the electronic design, assembly and testing. ISRO Satellite Centre (ISAC), Bengaluru provided the mechanical design, quality consultation and project management. The Inter University Centre for Astronomy and Astrophysics (IUCAA), Pune did the Coded Mask design, instrument calibration, and Payload Operation Centre. Space Application Centre (SAC) at Ahmedabad provided the analysis software. Physical Research Laboratory (PRL) Ahmedabad, provided the polarisation detection algorithm and ground calibration. A vast number of industries participated in the fabrication and the University sector pitched in by participating in the test and evaluation of the payload.

The Indian Space Research Organisation funded, managed and facilitated the project.

We thank the satellite and instrument teams for sharing their design details, and engineers at ISAC, ISITE, ISRO for providing CAD files that could be used in the mass model. In particular, we thank Prof. Shyam Tandon (IUCAA, Pune), Prof. H.M. Anitia (TIFR, Mumbai), Mr. Harshit Shah (TIFR, Mumbai), and Mr. Nagabhushana S. (IIA, Bangalore) for their assistance. We thank Mr. Dhanraj Borgaonkar (IUCAA, Pune) for the help in profiling the mass model. We thank Gaurav Waratkar (IIT Bombay) and  Vedant Shenoy (IIT Bombay) for assisting in the data analysis.

We acknowledge the use of Vikram-100 HPC at the Physical Research Laboratory (PRL), Ahmedabad and Pegasus HPC at the Inter University Centre for Astronomy and Astrophysics (IUCAA), Pune.

This work utilised various software including Python, AstroPy \citep{astropy}, NumPy \citep{numpy}, Matplotlib \citep{matplotlib}, IDL Astrolib \citep{landsman93}, FTOOLS \citep{blackburn95}, C, and C++.

\vspace{-1em}


\bibliography{main_biblio}


\appendix

\section{Visual impact of residual Poisson noise}\label{sec:vispoisson}
\begin{figure}[hbp]
  \centering
  \includegraphics[width=\linewidth]{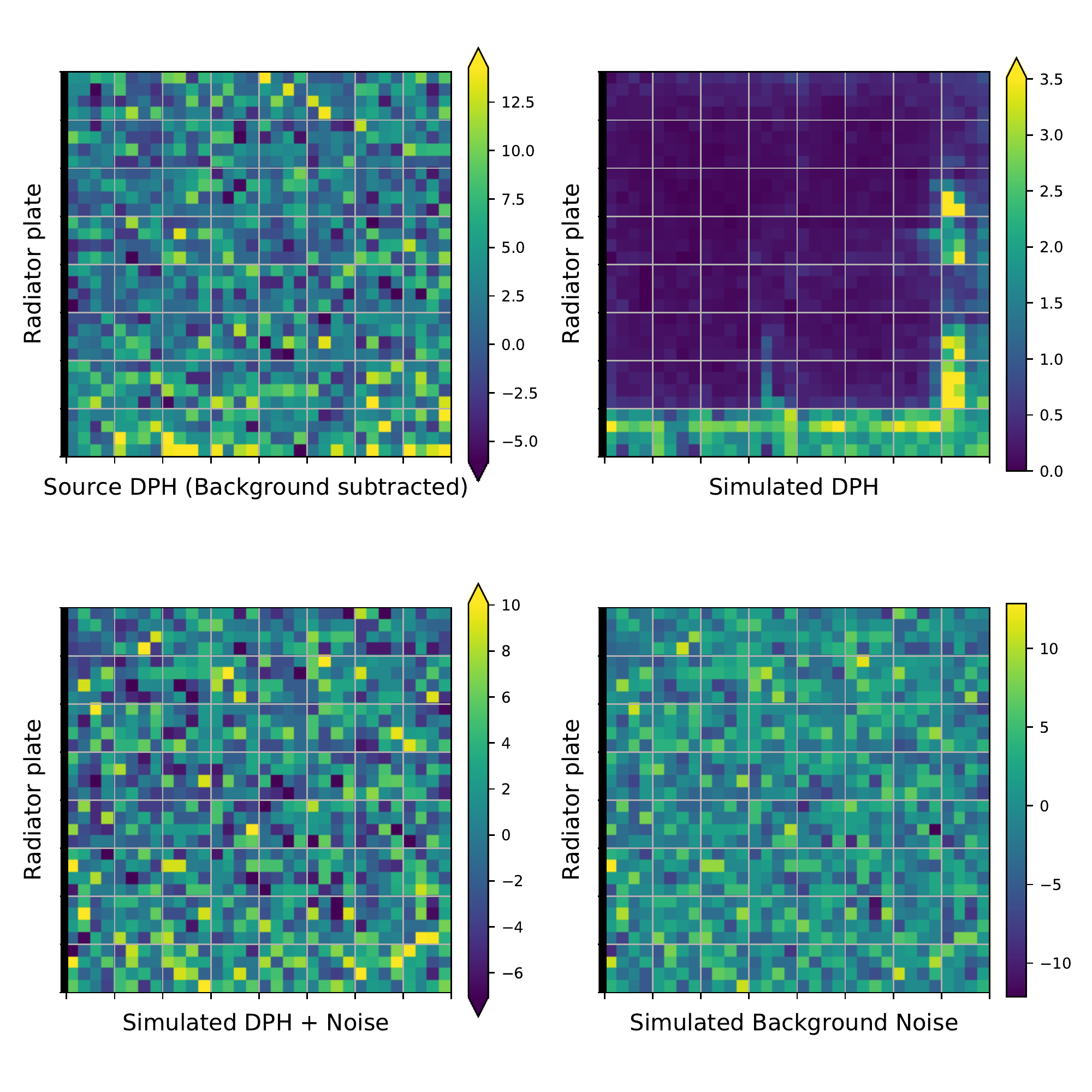}
  \caption{\textit{Upper left:} Background-subtracted source DPH for GRB~160607A. 
  \textit{Lower left:} Simulated DPH including a residual Poisson noise component. For both these panels, the colour bar ranges from first to 99th percentile values. 
  \textit{Upper right:} Simulated mass model DPH, with the maximum set at 99th percentile value.
  \textit{Lower right:} Simulated Poisson background noise, with the mean value subtracted. The range of the colour bar is the actual range for this residual Poisson noise.}
  \label{fig:GRB160607A_noise}
\end{figure}

\begin{figure}[hbtp]
  \centering
  \includegraphics[width=\linewidth]{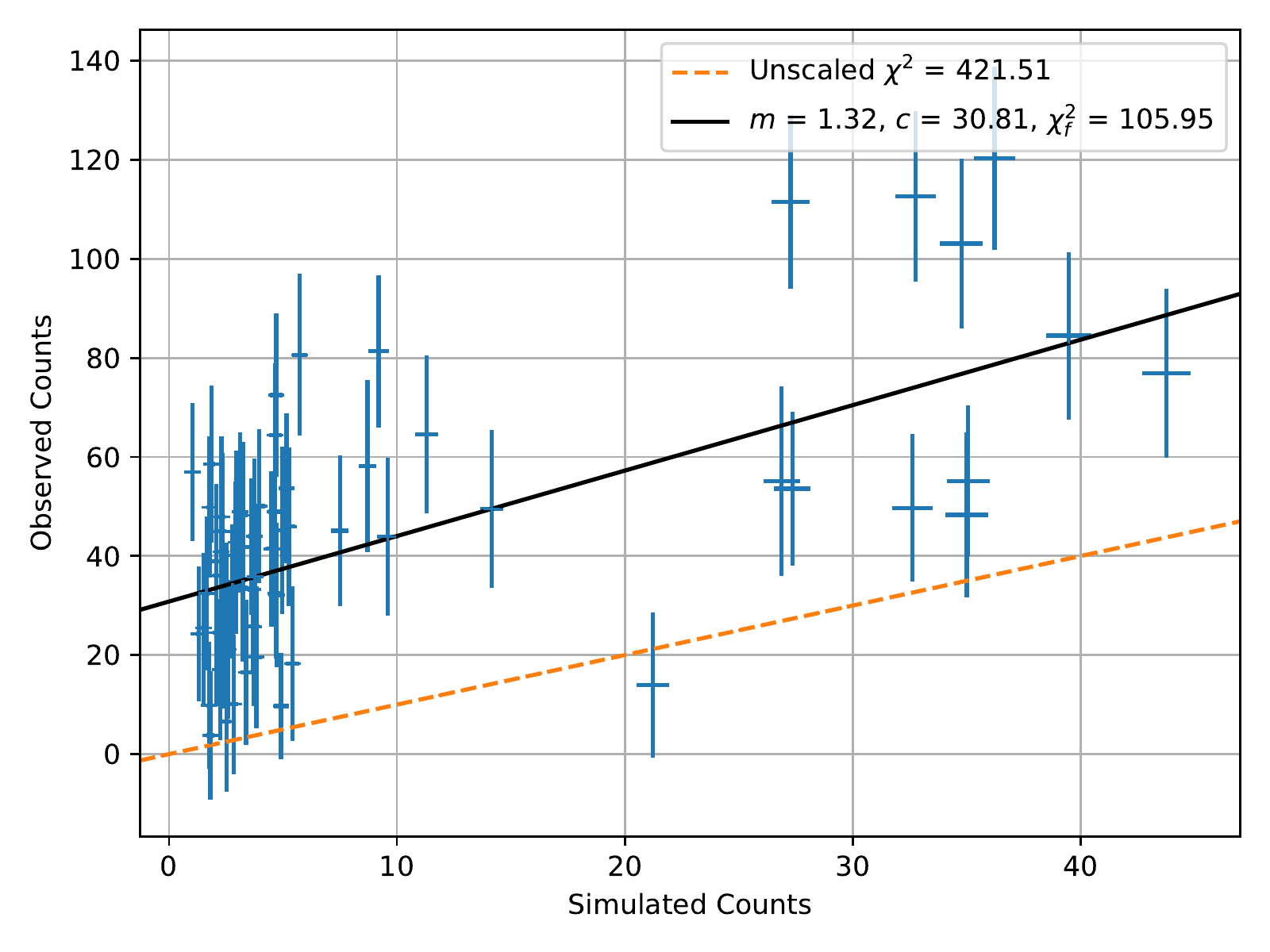}
  \caption{The observed and simulated module-wise count rates of GRB~160607 show a reasonable correlation. The best-fit line gives a scaling factor of 1.3}
  \label{fig:GRB160607A_s}
\end{figure}

For a few GRBs, we see that the scatter plots of module-wise observed versus simulated counts show a good correlation, but the DPHs are visually discrepant. A key factor in this is Poisson noise present in the observed data. We illustrate this with the example of GRB~160607A. In Figure~\ref{fig:GRB160607A_noise}, the upper left and right panels show the observed and simulated DPH for the GRB. These DPHs appear quite distinct. However, Figure~\ref{fig:GRB160607A_s} shows that there is a modest correlation between the observed and simulated module-wise count rates. In particular, the modules with high simulated counts ($\gtrsim 25$) also have high observed counts ($\gtrsim 40$). 

The key cause of this visual discrepancy is residual Poisson noise. For all of our GRBs, source DPH is obtained by creating a DPH for the GRB interval and subtracting a background DPH estimated from pre-- and post--GRB intervals. These intervals are selected to be much longer than the GRB, and the count rates are scaled to the GRB duration to suppress uncertainty in the estimate of the background. However, this removes only the mean background from the actual GRB data, leaving residual noise.

To demonstrate this, we create a simulated background DPH using a Poisson distribution with the expected background count rate. We then subtract the mean from this DPH to leave only the Poisson-induced noise (Figure~\ref{fig:GRB160607A_noise}, lower right panel). Adding this residual noise component to the simulation results in the DPH shown in the lower left panel: the prominent contrast from the simulation has been lost due to the high variations added by the residual noise. 

\section{Comparisons for the full GRB sample}\label{sec:othergrbs}
In \S\ref{sec:compare} and \S\ref{sec:quant} we have discussed the comparisons between the observed and simulated data for eight selected GRBs. Here, we discuss twenty more GRBs that were studied in this work.

\begin{figure*}[hp]
\centering

\begin{subfigure}{\textwidth}
  \centering
  \includegraphics[width=\linewidth]{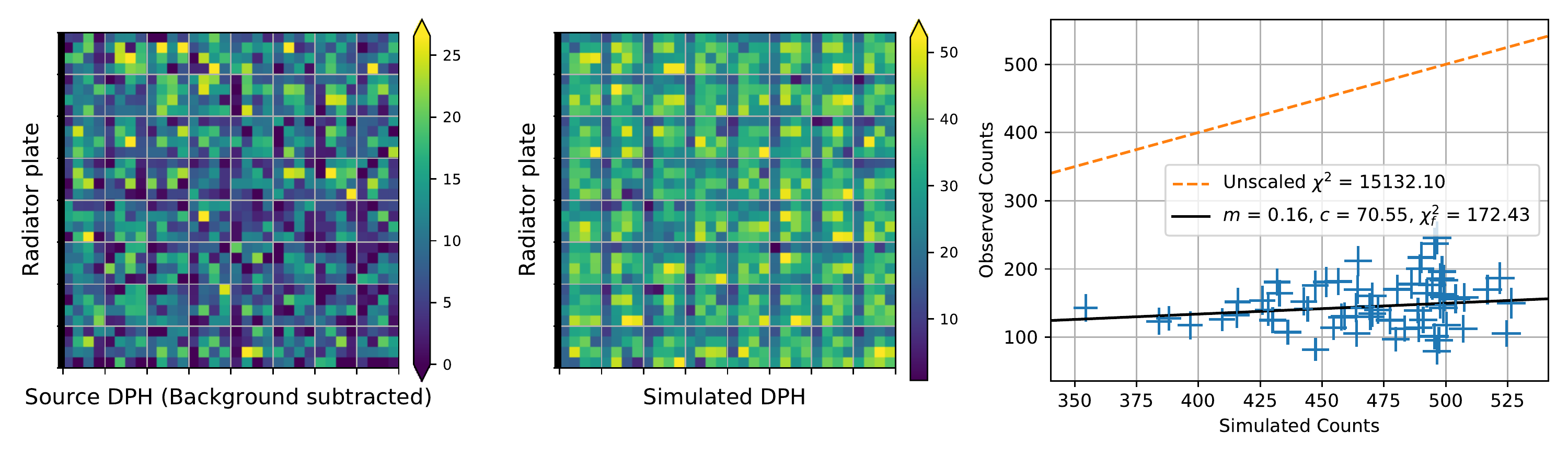}
  \caption{GRB160325A: $\theta=0.65\degr$, $\phi = 159.48\degr$}
  \label{fig:GRB160325A}
\end{subfigure}

\begin{subfigure}{\textwidth}
  \centering
  \includegraphics[width=\linewidth]{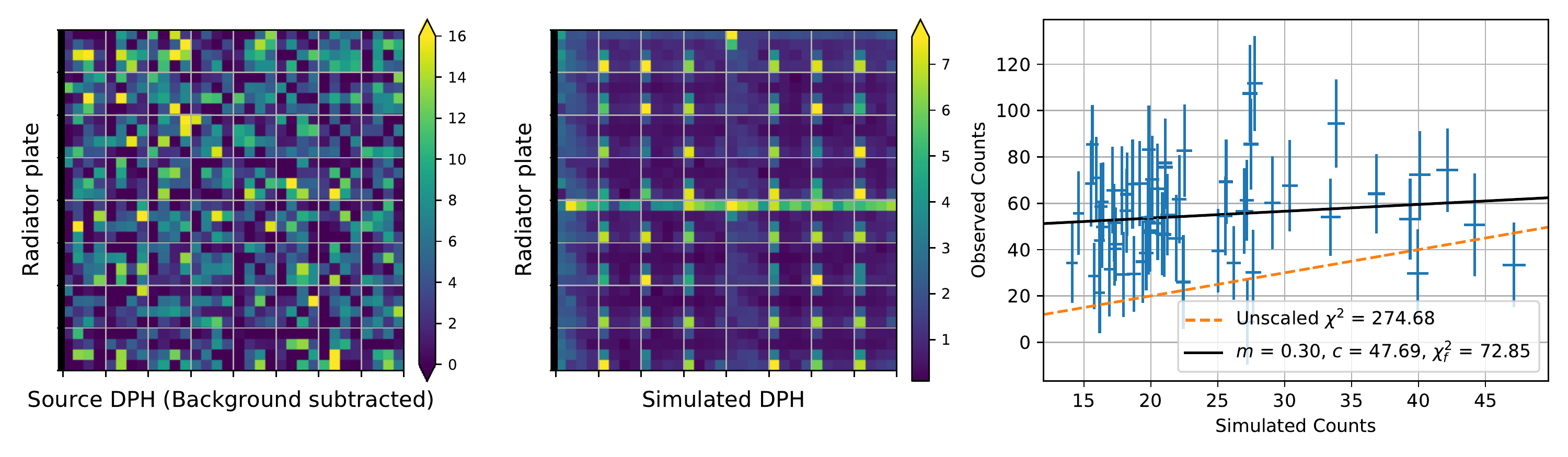}
  \caption{GRB170726A: $\theta=5.46\degr$, $\phi = 147.88\degr$}
  \label{fig:GRB170726A}
\end{subfigure}

\begin{subfigure}{\textwidth}
  \centering
  \includegraphics[width=\linewidth]{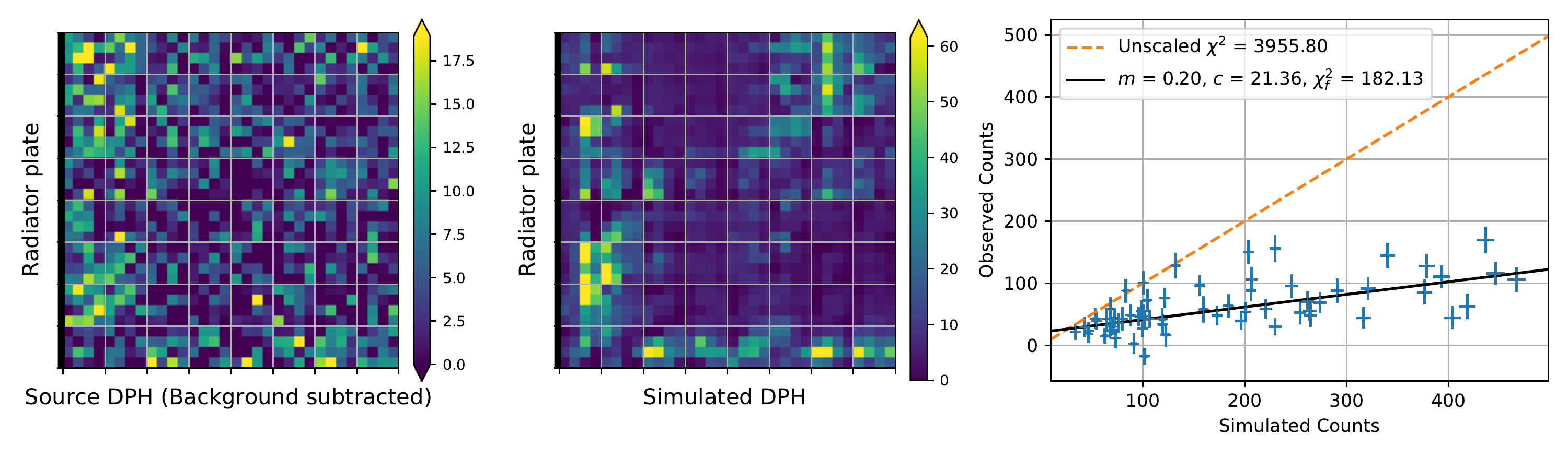}
  \caption{GRB180120A: $\theta=15.88\degr$, $\phi = 206.27\degr$}
  \label{fig:GRB180120A}
\end{subfigure}

\begin{subfigure}{\textwidth}
  \centering
  \includegraphics[width=\linewidth]{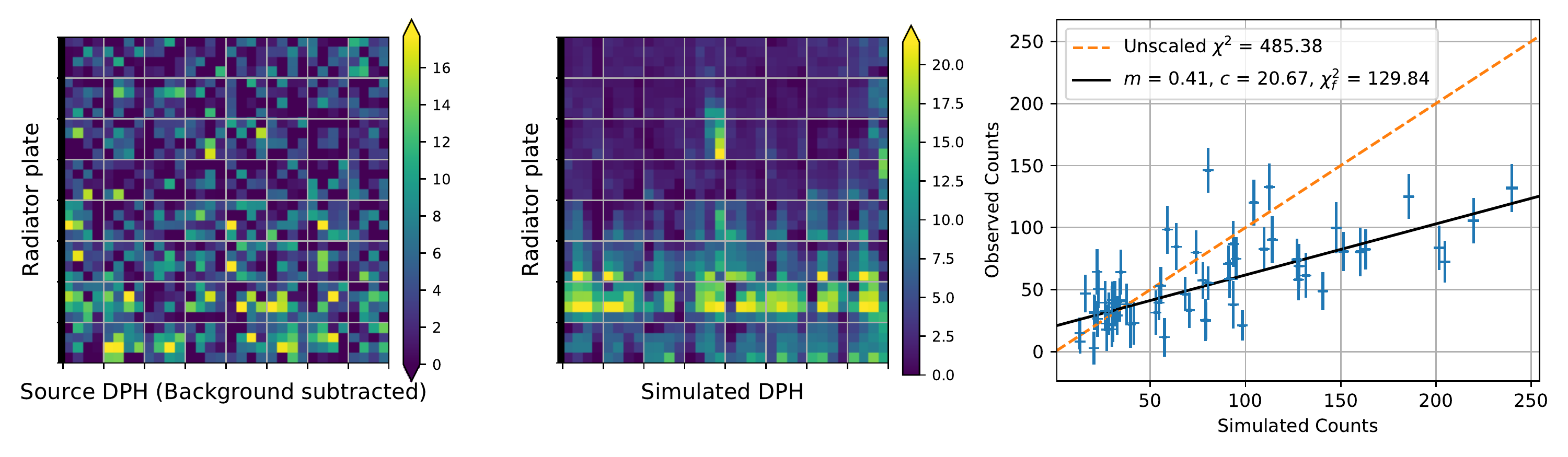}
  \caption{GRB180605A: $\theta=33.90\degr$, $\phi = 277.37\degr$}
  \label{fig:GRB180605A}
\end{subfigure}
\caption{Observed and simulated DPHs for GRBs incident from above the detector plane, $\theta \leq 60\degr$. Details are as in Figure~\ref{fig:grbplots1}.} \label{fig:grbplots3}
\end{figure*}

\begin{figure*}[hp]
\centering

\begin{subfigure}{\textwidth}
  \centering
  \includegraphics[width=\linewidth]{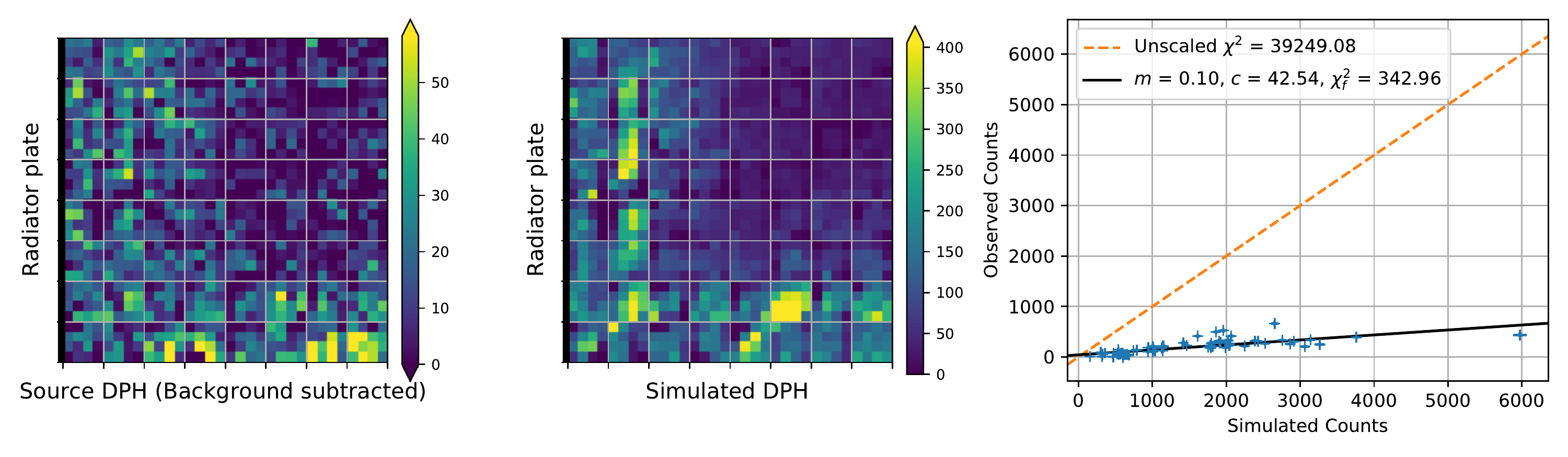}
  \caption{GRB180914A: $\theta=40.68\degr$, $\phi = 215.76\degr$}
  \label{fig:GRB180914A}
\end{subfigure}

\begin{subfigure}{\textwidth}
  \centering
  \includegraphics[width=\linewidth]{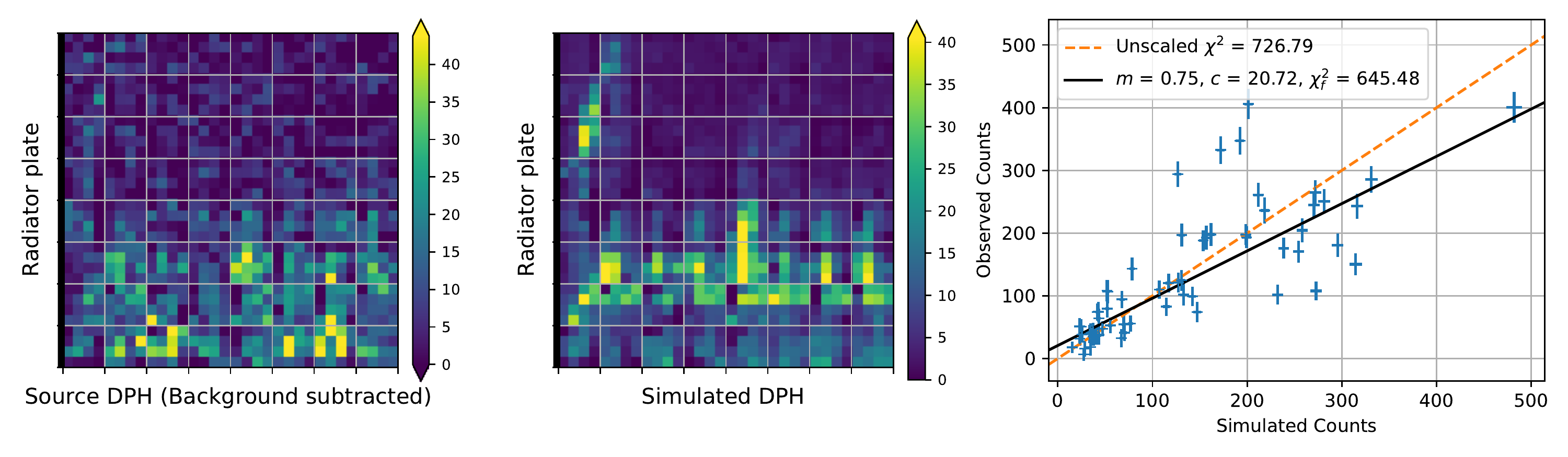}
  \caption{GRB180427A: $\theta=42.17\degr$, $\phi = 250.99\degr$}
  \label{fig:GRB180427A}
\end{subfigure}

\begin{subfigure}{\textwidth}
  \centering
  \includegraphics[width=\linewidth]{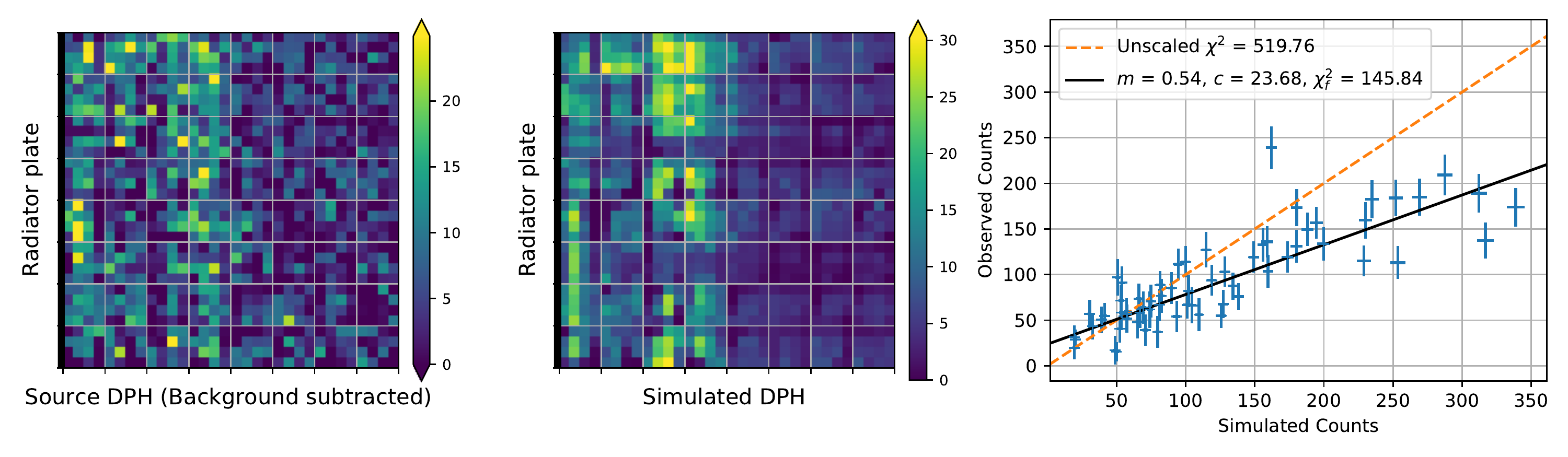}
  \caption{GRB190117A: $\theta=51.27\degr$, $\phi = 178.51\degr$}
  \label{fig:GRB190117A}
\end{subfigure}

\begin{subfigure}{\textwidth}
  \centering
  \includegraphics[width=\linewidth]{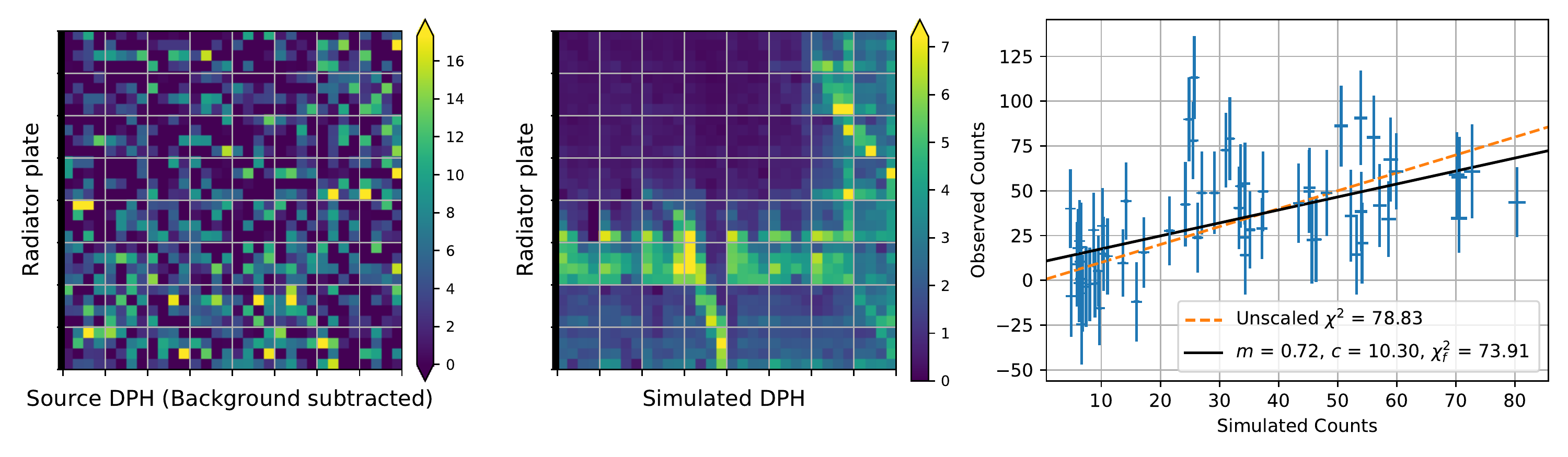}
  \caption{GRB170822A: $\theta=51.86\degr$, $\phi = 296.33\degr$}
  \label{fig:GRB170822A}
\end{subfigure}
\caption{Observed and simulated DPHs for GRBs incident from above the detector plane, $\theta \leq 60\degr$. Details are as in Figure~\ref{fig:grbplots1}.} \label{fig:grbplots4}
\end{figure*}

The observed and simulated DPHs of the on-axis GRB~160325A show good agreement, including vertical and horizontal patterns caused by shadows of the collimators (Figure~\ref{fig:GRB160325A}). The DPHs for GRB~170726A, located just outside the primary field of view, is more discrepant: but in reasonable agreement in terms of counts per detector (Figure~\ref{fig:GRB170726A}). 
In \S\ref{sec:compare} we had pointed out that the observed DPH of GRB~170527A shows two bright spots caused due to scattering from the alpha detector holders (Figure~\ref{fig:GRB170527A}), an effect we could not replicate in our simulations.
A similar effect is seen at the bottom of the observed DPH for GRB~171010A which is incident from $\phi=242\degr$ (Figure~\ref{fig:GRB171010A}), 
GRB~180605A with $\phi=277\degr$ (Figure~\ref{fig:GRB180605A}),
GRB~180914A with $\phi=217\degr$ (Figure~\ref{fig:GRB180914A}), 
and GRB~180427A with $\phi=251\degr$ (Figure~\ref{fig:GRB180427A}).

\begin{figure*}[hp]
\centering

\begin{subfigure}{\textwidth}
  \centering
  \includegraphics[width=\linewidth]{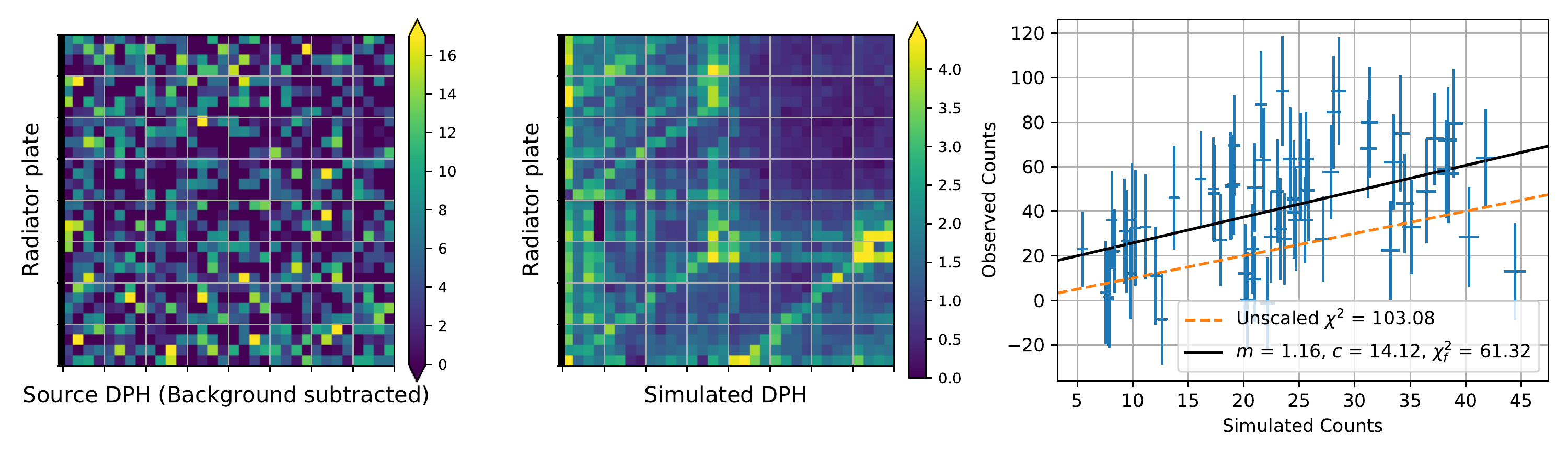}
  \caption{GRB171027A: $\theta=66.01\degr$, $\phi = 216.13\degr$}
  \label{fig:GRB171027A}
\end{subfigure}

\begin{subfigure}{\textwidth}
  \centering
  \includegraphics[width=\linewidth]{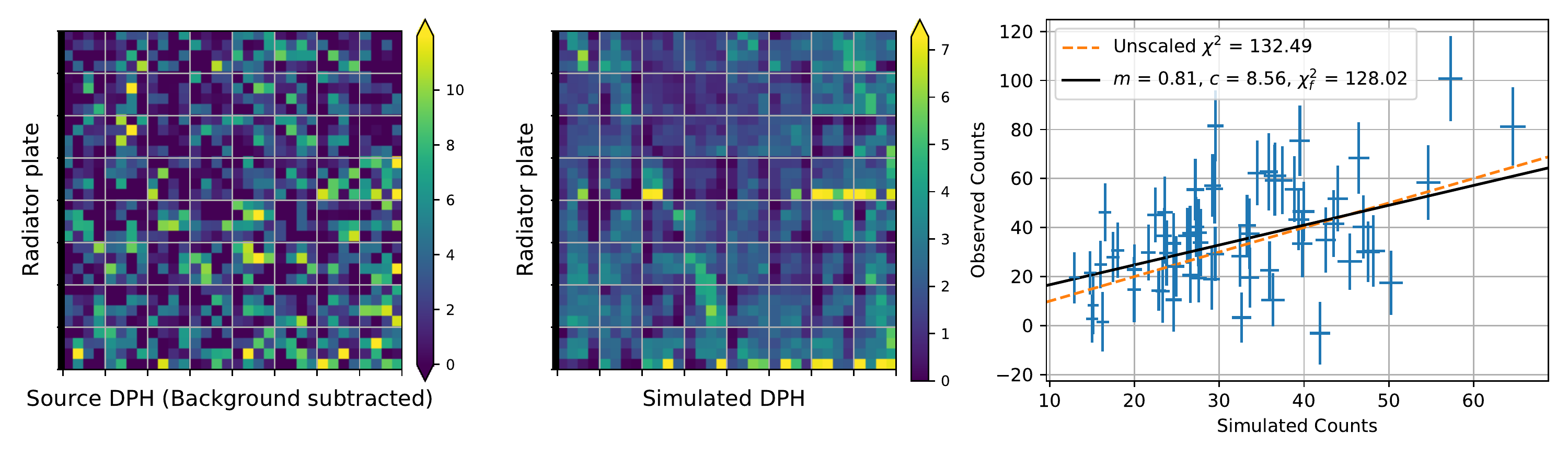}
  \caption{GRB190519A: $\theta=77.14\degr$, $\phi = 290.53\degr$}
  \label{fig:GRB190519A}
\end{subfigure}

\begin{subfigure}{\textwidth}
  \centering
  \includegraphics[width=\linewidth]{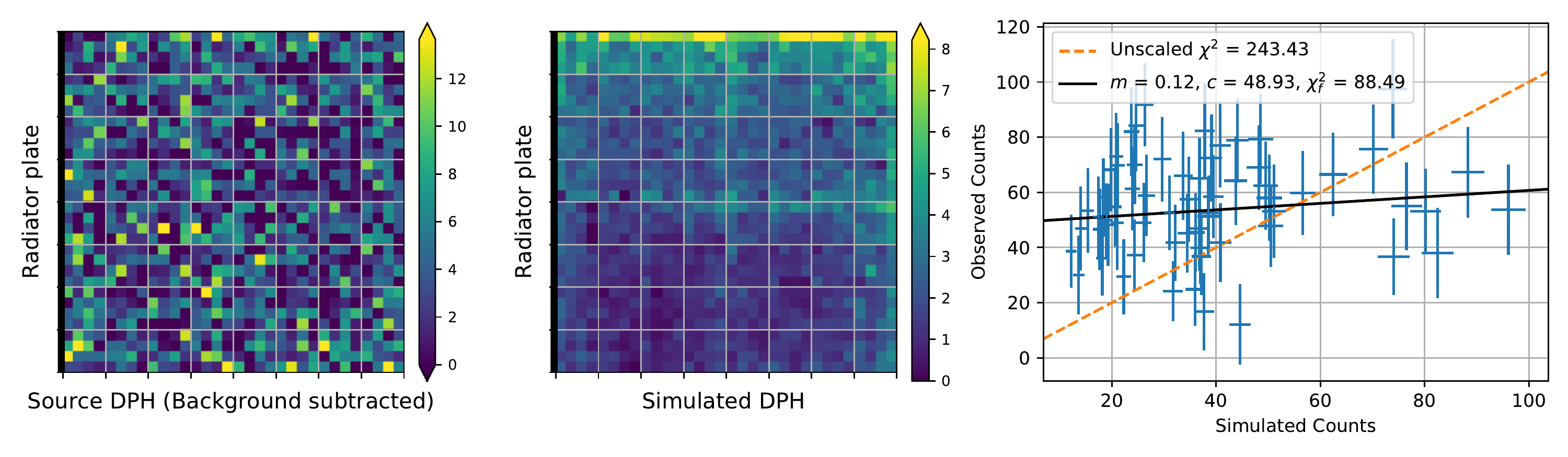}
  \caption{GRB160530A: $\theta=91.72\degr$, $\phi = 67.94\degr$}
  \label{fig:GRB160530A}
\end{subfigure}

\begin{subfigure}{\textwidth}
  \centering
  \includegraphics[width=\linewidth]{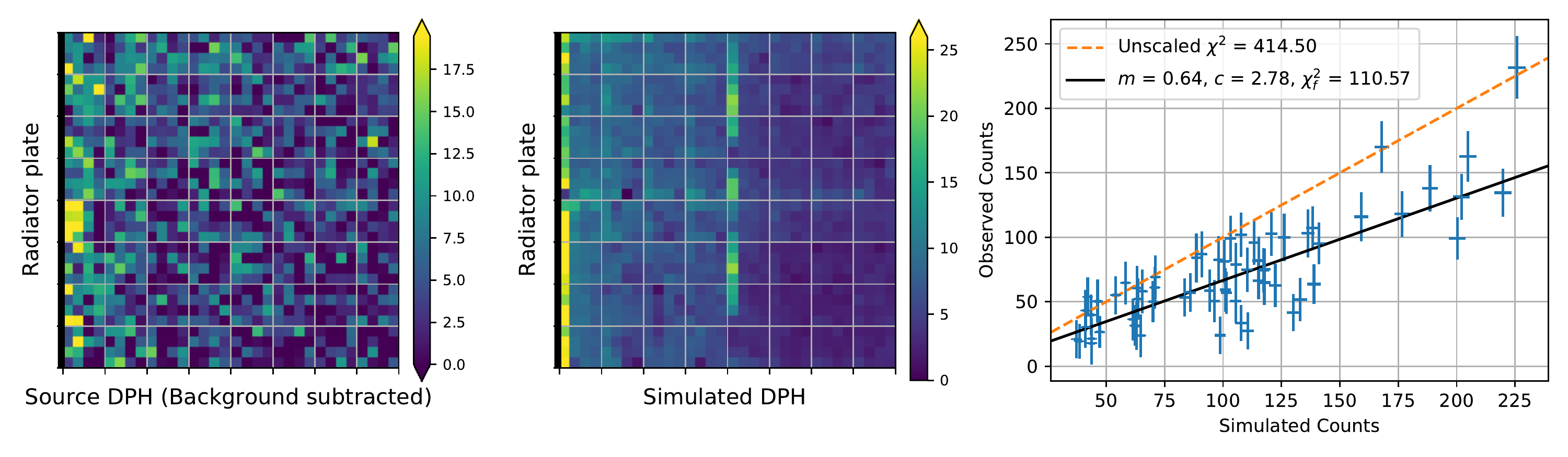}
  \caption{GRB170607B: $\theta=97.25\degr$, $\phi = 169.64\degr$}
  \label{fig:GRB170607B}
\end{subfigure}
\caption{Observed and simulated DPHs for GRBs incident at oblique angles, $60\degr < \theta \leq 120\degr$. Details are as in Figure~\ref{fig:grbplots1}.} \label{fig:grbplots5}
\end{figure*}
\begin{figure*}[hp]
\centering

\begin{subfigure}{\textwidth}
  \centering
  \includegraphics[width=\linewidth]{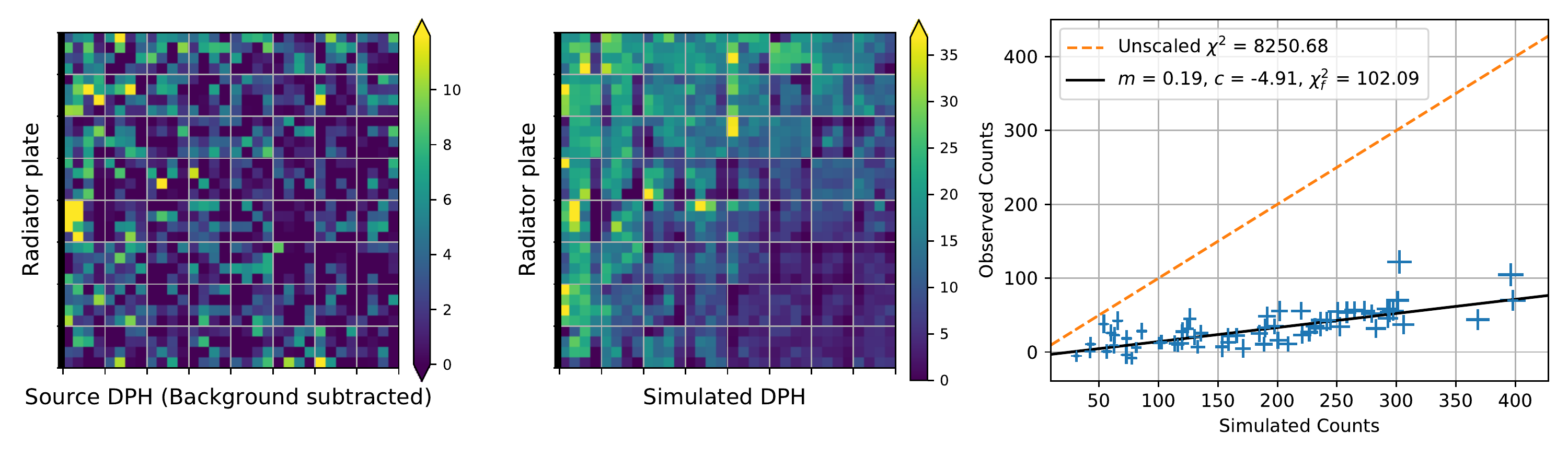}
  \caption{GRB180728A: $\theta=97.66\degr$, $\phi = 146.24\degr$}
  \label{fig:GRB180728A}
\end{subfigure}

\begin{subfigure}{\textwidth}
  \centering
  \includegraphics[width=\linewidth]{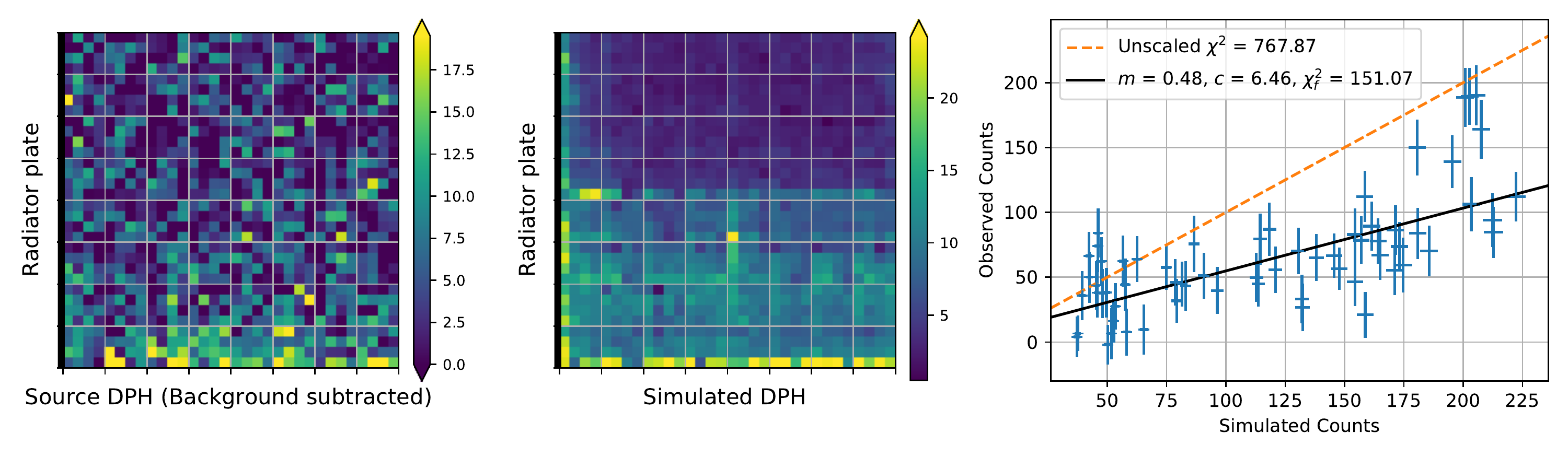}
  \caption{GRB160106A: $\theta=106.12\degr$, $\phi = 255.69\degr$}
  \label{fig:GRB160106A}
\end{subfigure}

\begin{subfigure}{\textwidth}
  \centering
  \includegraphics[width=\linewidth]{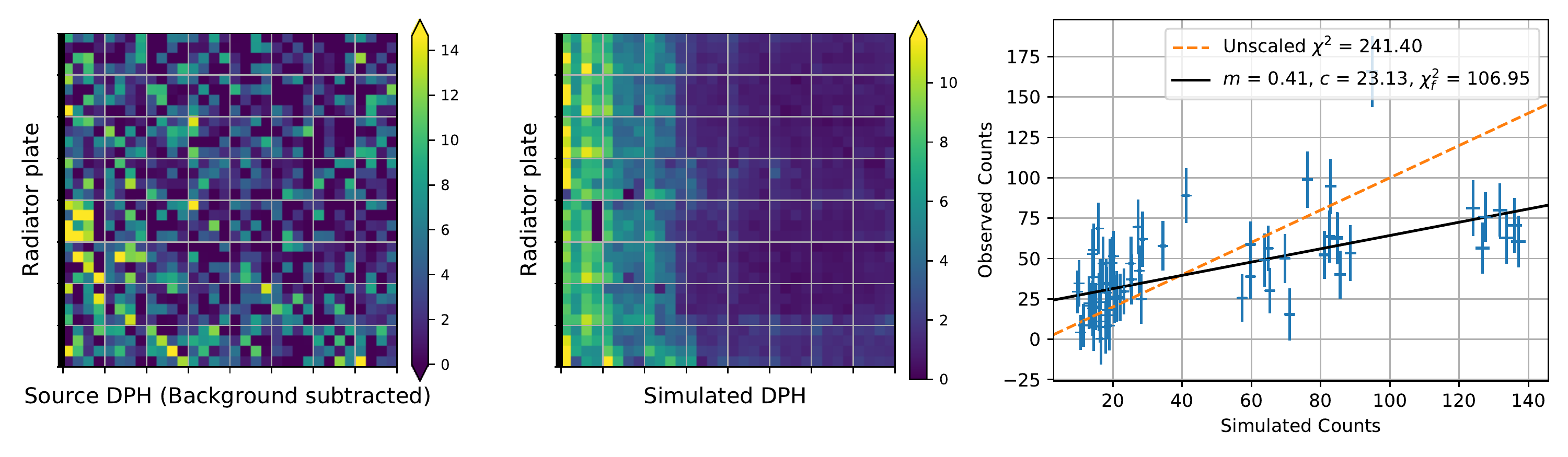}
  \caption{GRB170121B: $\theta=115.98\degr$, $\phi = 200.58\degr$}
  \label{fig:GRB170121B}
\end{subfigure}

\begin{subfigure}{\textwidth}
  \centering
  \includegraphics[width=\linewidth]{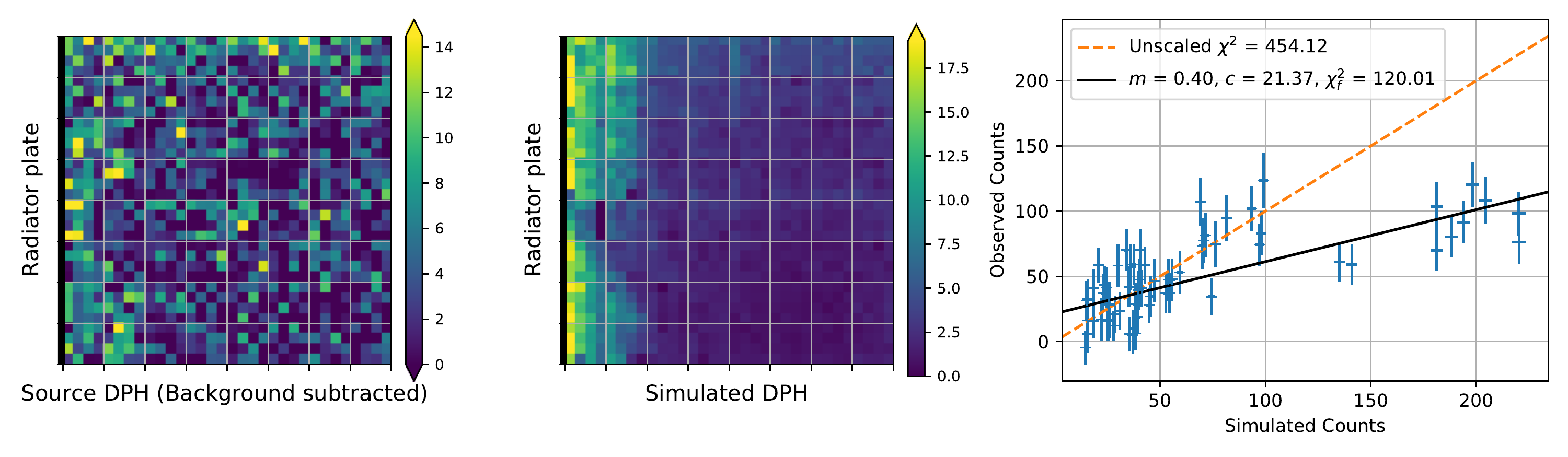}
  \caption{GRB170115B: $\theta=116.27\degr$, $\phi = 132.60\degr$}
  \label{fig:GRB170115B}
\end{subfigure}
\caption{Observed and simulated DPHs for GRBs incident at oblique angles, $60\degr < \theta \leq 120\degr$. Details are as in Figure~\ref{fig:grbplots1}.} \label{fig:grbplots6}
\end{figure*}

Figures~\ref{fig:grbplots5} \& \ref{fig:grbplots6} compare the observed and simulated DPHs for GRBs incident at oblique angles, $60\degr < \theta \leq 120\degr$. The observed DPHs are relatively featureless here, owing to the oblique angle of incidence. 
In \S\ref{sec:compare} we discussed GRB~170511A (Figure~\ref{fig:GRB170511A}) as an example of how the edge pixels of a quadrant can get significantly higher counts in such oblique cases: an effect slightly overestimated in our simulations.
This effect is also seen at the bottom edges of GRB~160106A (Figure~\ref{fig:GRB160106A}), and the top edge of GRB~160530A (Figure~\ref{fig:GRB160530A}). Even the visually discrepant GRB~171027A (Figure~\ref{fig:GRB171027A}) shows a good correlation in the module-wise count rates, with slope close to unity. Such visual discrepancy is discussed in \ref{sec:vispoisson}.

\begin{figure*}[hp]
\centering

\begin{subfigure}{\textwidth}
  \centering
  \includegraphics[width=\linewidth]{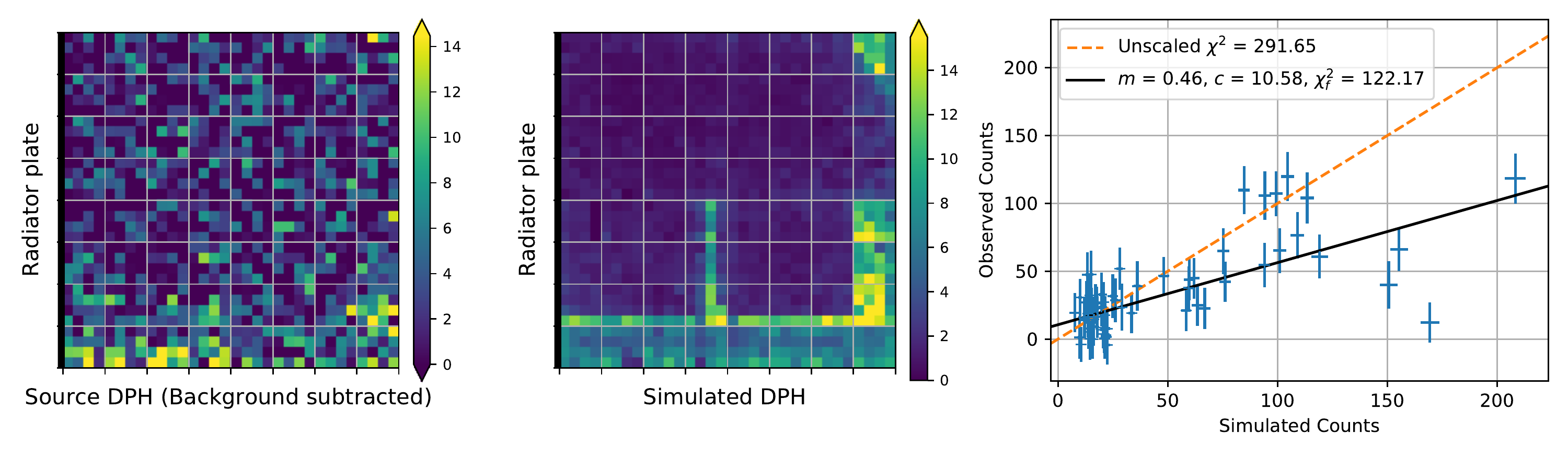}
  \caption{GRB170921B: $\theta=136.68\degr$, $\phi = 302.73\degr$}
  \label{fig:GRB170921B}
\end{subfigure}

\begin{subfigure}{\textwidth}
  \centering
  \includegraphics[width=\linewidth]{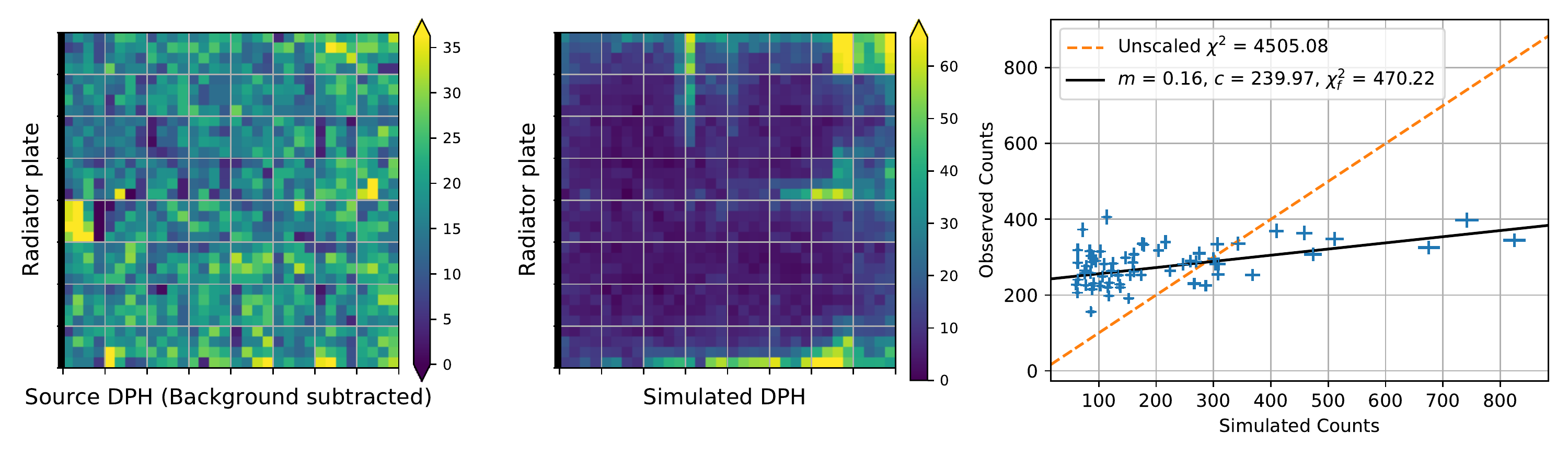}
  \caption{GRB170614A: $\theta=137.69\degr$, $\phi = 340.67\degr$}
  \label{fig:GRB170614A}
\end{subfigure}

\begin{subfigure}{\textwidth}
  \centering
  \includegraphics[width=\linewidth]{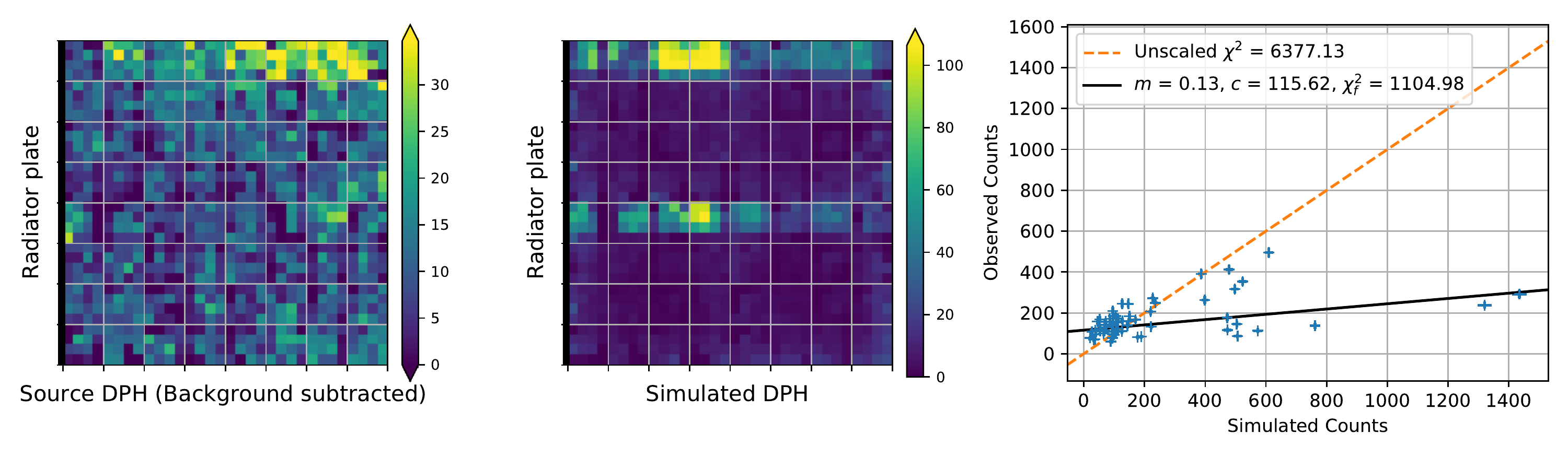}
  \caption{GRB190530A: $\theta=154.50\degr$, $\phi = 80.31\degr$}
  \label{fig:GRB190530A}
\end{subfigure}

\begin{subfigure}{\textwidth}
  \centering
  \includegraphics[width=\linewidth]{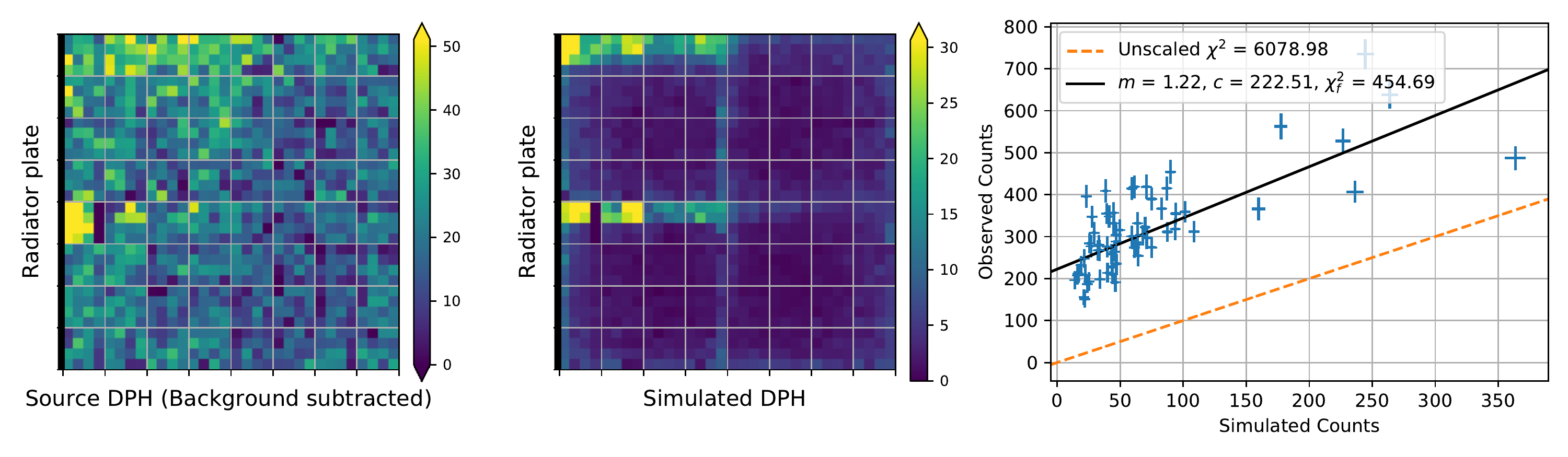}
  \caption{GRB160821A: $\theta=156.18\degr$, $\phi = 59.27\degr$}
  \label{fig:GRB160821A}
\end{subfigure}
\caption{Observed and simulated DPHs for GRBs incident from below the detector plane, $\theta > 120\degr$. Details are as in Figure~\ref{fig:grbplots1}.} \label{fig:grbplots7}
\end{figure*}

Figure~\ref{fig:grbplots7} shows comparisons for GRBs incident from below the focal plane ($\theta > 120\degr$). Here we see greater discrepancies between observations and simulations. This is an unsurprising effect, as the satellite body has not been modelled very accurately. 
As discussed in \S\ref{sec:compare}, a common discrepancy seen here is the presence of ``hotspots'' in the simulation: a certain region of the DPH is disproportionately brighter than the rest. Such an effect is seen in the top right modules of GRB~170614A (Figure~\ref{fig:GRB170614A}) and two detector modules along the top row of GRB~190530A (Figure~\ref{fig:GRB190530A}).
There is still broad agreement in the observations and simulations: the bright lower edge, right side, and middle ``spike'' in the simulations of GRB~170921B can be discerned in the observed data (Figure~\ref{fig:GRB170921B}).
Overall, it appears as if the observed DPHs are ``blurred'' versions of the simulations, the sharper simulated DPHs are likely an artefact of our choice of clumping several satellite components into compact boxes and sheets.

\balance

\end{document}